\documentclass[%
 reprint,
 amsmath,amssymb,
 aps,
]{revtex4-1}

\usepackage{graphicx}% Include figure files
\usepackage{dcolumn}% Align table columns on decimal point
\usepackage{bm}% bold math
\usepackage{subfig}
\begin{document}

%\preprint{APS/123-QED}

\title{Active transport in a channel: stabilisation by flow or thermodynamics}% Force line breaks with \\

\author{Santhan Chandragiri}
 \altaffiliation{Department of Chemical Engineering, Indian Institute of Technology Madras, Chennai 600036, India. E-mail: sumesh@iitm.ac.in}%Lines break automatically or can be forced with \\
\author{Amin Doostmohammadi}%
%\affiliation{%
\author{Julia M Yeomans}
\altaffiliation{The Rudolf Peierls Centre for Theoretical Physics, Clarendon Laboratory, Parks Road, Oxford, OX1 3PU,UK. E-mail: julia.yeomans@physics.ox.ac.uk}
\author{Sumesh P Thampi}


%\affiliation{
%\affiliation{
 %Third institution, the second for Charlie Author
%\affiliation{%



\begin{abstract}
Recent experiments on active materials, such as dense bacterial suspensions and microtubule-kinesin motor mixtures, show a promising potential for achieving self-sustained flows. However, to develop active microfluidics it is necessary to understand the behaviour of active systems confined to channels. Therefore here we use continuum simulations to investigate the behaviour of  active fluids in a two-dimensional channel. Motivated by the fact that most experimental systems show no ordering in the absence of activity, we concentrate on temperatures where there is no nematic order in the passive system, so that any nematic order is induced by the active flow. We systematically analyze the results, identify several different stable flow states, provide a phase diagram and show that the key parameters controlling the flow are the ratio of channel width to the length scale of active flow vortices, and whether the system is flow aligning or flow tumbling.
%\begin{description}
%use the optional argument of the \verb+\item+ command to give the category of %each item. 
%\end{description}
\end{abstract}

\pacs{Valid PACS appear here}% PACS, the Physics and Astronomy
\maketitle

%\tableofcontents

\section{Introduction}

Collective behaviour is present in biological systems at all length scales. Examples include cytoskeletal elements \cite{Kruse2004, Schaller2010}, cell colonies \cite{Saw2017,Duclos2018}, bacterial suspensions \cite{Wensink2012,Dunkel2013}, locust swarms \cite{Buhl2006} and fish schools \cite{Lopez2012}. Similar behaviour is seen in  synthetic systems, such as catalytic colloids \cite{Harder2018,Buttinoni2013} and vibrated granular rods \cite{Narayan2007}. These, active, materials are driven out of equilibrium by continuous energy input at the level of single particles. Despite the differences in the origin of their activity, several features such as long-ranged ordering patterns, hydrodynamic instabilities and turbulent-like flow dynamics are prevalent among different active systems\cite{Marchetti2013,Koch2011, Buttinoni2013, Doostmohammadi2018, Ramaswamy2010,Opathalage2018}. The extent and origins of this universality is not yet fully understood, and controlling the pattern formation and flow dynamics is a way to gain insight into the underlying physics that governs the behaviour of active matter.

Though active systems have been a subject of increasingly intense study in the past decade, limited progress has been made towards utilising them for real-life applications. One of the major hurdles is in controlling their flow behaviour, but manipulation of the geometry confining the active fluid may be a way forward. For example, a net mass flux through rectangular channels containing microtubule bundles powered by kinesin motors has been achieved by tuning the aspect ratio of the channels \cite{Wu2017}, with potentials in microfluidics applications.

 In this article we use continuum simulations to investigate the behaviour of active fluids confined in a two-dimensional channel. We consider active nematics\cite{Doostmohammadi2018}, a class of active fluids such as microtubule bundles\cite{Sanchez2012}, elongated bacteria \cite{Volfson2008, Doostmohammadi2016, Shi2014} and fibroblast\cite{Kemkemer2000,Duclos2014}, epithelial\cite{Saw2017}, and neural progenitor stem cells\cite{Kawaguchi2017}, that manifest nematic liquid crystal features including nematic order and topological defects. These have been shown to undergo a spontaneous transition to a flowing state at a finite activity\cite{Voituriez2005,Duclos2018}, and then to a `dancing' state at higher activities~\cite{Shendruk2017}, where velocity vortices drive topological defects along oscillatory paths. Further increase in activity drives the system to an active turbulent state where a chaotic flow regime is accompanied by the continuous creation, movement and annihilation of   $+1/2$ and $-1/2$ topological defects. The rate of defect creation and annihilation balance such as to maintain a statistically steady-state defect density \cite{Sanchez2012,Thampi2013}. The defects are not simply advected, they have their own dynamics: $+1/2$ defects continuously move due to their innate self-propulsion \cite{Giomi2013}, and both  $+1/2$ and $-1/2$  defects are strong sources of active flows \cite{Thampi2014}.

However, results so far have been for a system where a thermodynamic free energy imposes strong nematic ordering. For example, spontaneous flow transitions in confined active nematics \cite{Marchetti2013,Ramaswamy2010} and emergence of various flow states such as dancing state~\cite{Shendruk2017} and its transition to turbulence~\cite{Doostmohammadi2017} all assume the presence of nematic order in the system in the absence of activity.
Many experimental systems have no nematic ordering in the absence of activity. We therefore consider temperatures above the nematic transition. We argue that local nematic order can be generated by an active flow field, even in the absence of thermodynamic ordering in such systems, and describe the flow configurations of such a flow-stabilised system in a channel. By comparing results for thermodynamically-stabilised and flow-stabilised active nematics we show that the key parameters controlling the flow states are the activity, the channel width and whether the active system is flow aligning or flow tumbling.

We describe our governing equations and simulation  approach in Sec.~\ref{sec:simu_method}. We then summarise how active nematics can become ordered both below and above the nematic transition temperature. The possible flow configurations are characterised in Sec.~\ref{sec:fourstates}. In Sec.~ \ref{sec:results} we present phase plots showing the different flow states that are stabilised as activity, the flow alignment parameter and the temperature are varied.

% We describe the flow states of an active paranematic when confined in a channel in section\ref{sec:isonematic}. In sections \ref{sec:fourstates} - \ref{sec:results} we discuss our results on the role of temperature and show the decisive and unifying role of an \textit{effective} aligning parameter in obtaining different flow states. Discussion of results and conclusion are followed then. 

%\ju{to be rewritten at end eg doesn't yet mention that we are looking at confined systems}

%%%%%%%%%%%%%%%%%%%%%%%%%%%%%%%%%%%%%%%%%%%%%%%%%%%%%%%%%%%%%%%%%%%%%%%%%%%%%
       
\section{Equations of Motion} 
\label{sec:simu_method}
We first summarise the continuum description of an active nematic, and the simulation approach we use to solve the equations of motion. In particular we discuss how extensile flow fields can lead to local order in an active nematic even in the absence of thermodynamic ordering.
\subsection{Governing equations}
\label{sec:govn_eqns}
We consider the dynamics of an incompressible, active fluid of density $\rho$ and velocity $\mathbf{u}$ within the continuum approximation. The fluid is driven by active constituents which are assumed to be apolar in nature \cite{Marchetti2013}. We define a tensor order parameter 
\begin{equation} \label{eqn:Q}
Q_{ij} = \frac{q}{2}(3n_{i}n_{j}-\delta_{ij})
\end{equation}
where $\mathbf{n}$ is the director field (of unit length) and $q$ gives the strength of nematic ordering of the active constituents \cite{P1995}. 
The free energy of the system is defined using a Landau-de Gennes approach in the same way as for a passive nematic liquid crystal. It consists of two contributions, a bulk term
%
%\begin{subequations} 
\begin{align}
F_{\textnormal{LDG}} =\frac{A}{2}Q_{ij}Q_{ji}+\frac{B}{3}Q_{ij}Q_{jk}Q_{ki}+\frac{C}{4}\left(Q_{ij}Q_{ji}\right)^2,
%{\color{LimeGreen}A = \frac{A_0\left(T - T_0\right)}{2};\hspace{0.2cm}T = T^*+ 27\beta(T_c-T^*)}
\label{eqn:LDG}
\end{align}
%\end{subequations}
%
where the coefficients $A, B, C$ determine the order in the system, and  
a gradient term modelled using the Frank-Oseen free energy 
\begin{equation} \label{eqn:FO}
F_{\textnormal{FO}} = \frac{K}{2} \left(\partial_k Q_{ij}\right)^2 
\end{equation}
which describes the free energy cost of any deviations from uniform nematic ordering.
In eqn~(\ref{eqn:FO}) we have assumed a single elastic constant, $K$.  The total free energy of the active nematic is $F = F_{\textnormal{LDG}} + F_{\textnormal{FO}}$.

The mass and momentum conservation equations for the fluid are\cite{Yeomans2016,Beris1994}
%
%\textbf{Equation of continuity} 
\begin{equation} \label{eqn:mass}
\partial_iu_i = 0,
\end{equation}
%
%\textbf{Equation of motion} \label{eqn:mom}
%
%\begin{subequations}
%\begin{align}
\begin{equation}
\rho\left(\partial_t+u_k\partial_k \right)u_i  = \partial_j\pi_{ij} 
\end{equation}
where the stress tensor
\begin{equation}
\pi_{ij}  = \pi_{ij}^{\textnormal{viscous}} + \pi_{ij}^{\textnormal{active}} + \pi_{ij}^{\textnormal{passive}}.
\end{equation}
%\end{align}
%\end{subequations}
%
The viscous stress is modelled using a Newtonian constitutive relation,\cite{Batchelor2000}
\begin{equation} \label{eqn:viscous}
\pi_{ij}^{\textnormal{viscous}} = 2 \mu E_{ij} 
\end{equation}
where $\mu$ is the shear viscosity of the fluid and $\mathbf{E}$ is the symmetric part of the velocity gradient tensor. 

The active stress is taken to be proportional to the orientational order \cite{Simha2002}
\begin{equation} 
\pi_{ij}^{\textnormal{active}} =-\zeta Q_{ij} \label{eqn:active}
\end{equation}
where $\zeta$ describes the strength of the activity. %\ad{\sout{In a fluid with no orientational order, $\mathbf{Q}$ is proportional to $\mathbf{I}$. Therefore the active stress will just contribute to the pressure field.} [Note to Santhan: with no order q=0 thus Q=0. In any case, even for finite order, active stress has no contribution to pressure because the Q tensor is traceless]} 
When local order is developed in the fluid, either due to activity itself or due to lowering the temperature, this term can drive active flows. The sign of the activity coefficient $\zeta$ determines the symmetry of the stresslet flow field generated by the active constituents \cite{Simha2002}. $\zeta>0$ represents extensile systems which draw fluids in from their sides and push it out along the director field, e.g. bundled microtubules driven by motor proteins \cite{Edwards2009,Ramaswamy2010}. Reversing the direction of the active stress gives rise to contractile systems, $\zeta<0$. 

The  elasticity due to orientational order in the fluid generates stresses, that also arise in a passive liquid crystal. These are given by \cite{P1995,Beris1994}
\begin{equation} \label{eqn:passive}
\begin{split}
\pi_{ij}^{\textnormal{passive}} = -P\delta_{ij}+2 \lambda \left(Q_{ij}+\frac{\delta_{ij}}{3} \right) \left(Q_{kl}H_{lk} \right)-\lambda H_{ik} \left(Q_{kj}+\frac{\delta_{kj}}{3}\right)\\-\lambda \left(Q_{ik}+\frac{\delta_{ik}}{3}\right)H_{kj}-\partial_iQ_{kl}\left(\frac{\delta F} {\delta\partial_{j}Q_{lk}}\right)+Q_{ik}H_{kj}-H_{ik}Q_{kj} 
\end{split}
\end{equation}
where $\lambda$ is the aligning parameter 
%\ju{bad style to refer to eqn in the future} (see eqn~\ref{eqn:corot} also)
and $H_{ij}$ is the molecular potential field calculated from the free energy, $F$,
\begin{equation} \label{eqn:potential}
H_{ij} = -\frac{\partial F}{\partial Q_{ij}}+\frac{\delta_{ij}}{3}\textnormal{Tr}\left(\frac{\partial F}{\partial Q_{kl}}\right).
\end{equation}

Both the active and the passive stresses are dependent on the order parameter $\mathbf{Q}$. Any gradient in $\mathbf{Q}$, either in the orientation ($\mathbf{n}$) or strength ($q$), can result in a fluid flow. Moreover the fluid flow can, in turn, change the orientational order in the system. The dynamics of $\mathbf{Q}$ follows the evolution equation%\\ \textbf{Evolution equation of $\mathbf{Q}$}
\begin{equation} \label{eqn:evolution}
\left(\partial_t+u_k\partial_k\right)Q_{ij}-S_{ij} = \Gamma H_{ij}
\end{equation}
where $\Gamma$ is the orientational diffusivity and $S_{ij}$ is the generalized advection term
\begin{equation} \label{eqn:corot}
\begin{split}
S_{ij} = \left(\lambda E_{ik}+\Omega_{ik}\right)\left(Q_{kj}+\frac{\delta_{kj}}{3}\right)+\left(Q_{ik}+\frac{\delta_{ik}}{3}\right)\left(\lambda E_{kj}-\Omega_{kj}\right)\\-2\lambda\left(Q_{ij}+\frac{\delta_{ij}}{3}\right)\left(Q_{kl}\partial_ku_l\right)
\end{split}
\end{equation}
with $\boldsymbol{\Omega}$ the antisymmetric part of the velocity gradient tensor. 
$S_{ij}$ describes the response of the fluid to spatial gradients in the velocity field. 
Equation~\ref{eqn:corot} indicates that the aligning parameter $\lambda$ determines the relative importance of the response of $\mathbf{Q}$ to the extensional and rotational parts of the flow field.

\subsection{Simulation details}\label{sec:sims}

The equations of motion summarised in Sec.~\ref{sec:govn_eqns} are solved using a hybrid lattice Boltzmann algorithm. In this method, the fluid dynamics is obtained by solving a discretised Boltzmann equation which yields the Navier-Stokes equations in the continuum limit \cite{Denniston2002}. The evolution of the $\mathbf{Q}$ tensor is followed using the method of lines wherein the spatial derivatives are discretised using second order central differences and the temporal integration of the resulting algebraic equations is performed using a fourth order Runge-Kutta method \cite{Doostmohammadi2016}.

\begin{figure} 
	\centering
		\includegraphics[width=0.925\columnwidth]{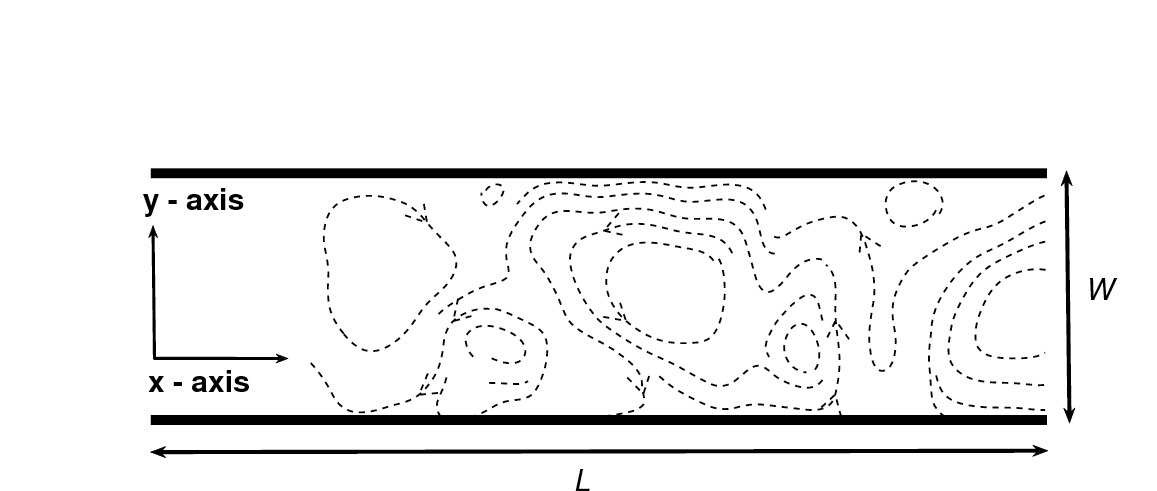} %\hspace{-0.6cm} 
%         \rotatebox{90} {\small{$W$}}
%        \\
%         \fbox{\small{channel length(L)}}
		\caption{Schematic of the simulation setup. The active fluid is confined between two rigid plates of length $L$ separated by a distance $W$. Dashed lines indicate typical streamlines corresponding to an active turbulent flow configuration.}
        \label{fig:chnl_schematic}
\end{figure}

A schematic of the two-dimensional domain used for the simulations is shown in Fig.~\ref{fig:chnl_schematic}. The active fluid is sandwiched between two parallel plates of length $L$ separated by a distance $W$. Typically $L >> W$ so that the results are independent of the channel length.
To facilitate the discussions, we adopt Cartesian coordinates, $x$ along the channel length, and $y$ in the perpendicular direction, across the channel. Periodic boundary conditions for the flow and orientation tensor are  imposed along the $x$ direction. No-slip boundary conditions are used for the velocity field at the walls. Homogeneous boundary conditions are used to align the director field parallel to the walls with a magnitude equal to that of the passive liquid crystal at the chosen simulation temperature. Hence $\mathbf{Q} = 0$ on the walls for temperatures above the passive nematic ordering temperature. %\sout{\ad{and in this regime the results are independent of the choice of homogeneous or homeotropic boundary conditions for $\mathbf{Q}$}. \ad{[is this correct? we had the results for homeotropic BC previously, maybe comment on that; also results in the ceilidh dance paper are for hometropic]}}
%\spt{AD-It is true that we get different flow states with homeotropic BC, but we don't have a phase diagram like fig9 for homeotropic yet - so it may be safer just not to say it before we simulate the effect of homeotropic BCs.}
To  mimic the experimental conditions, all the simulations are started with a quiescent fluid and with a random initial configuration for the orientation tensor.
The range of parameters used in the simulations are listed in Table \ref{tbl:paramters}.
\begin{table}
\small
  \caption{Parameters used in the simulations}
  \label{tbl:paramters}
  \begin{tabular*}{0.48\textwidth}{@{\extracolsep{\fill}}ccl}
    \hline
    Parameter & Range \\
    \hline
    $\lambda$ & 0.3 - 1.0 \\
    A & -0.05 - 0.015 \\
    B & 0 - 0.3 \\
    C & 0 - 0.3 \\
    $\zeta$  & 0.00015 - 0.07 \\
    $\Gamma$ & 0.34 \\
    $\mu$ & 0.667 \\
    $\rho$ & 1\\
    K & 0.01 \\
    $L$ & 100 - 400 \\
    $W$ & 20 - 70 \\
    \hline
  \end{tabular*}
\end{table}

\subsection{Mechanisms for nematic ordering} \label{sec:isonematic}
\begin{figure}
\centering
\includegraphics[width=\linewidth]{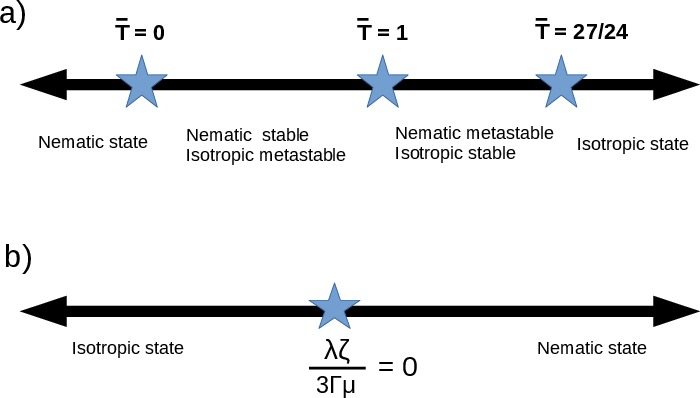}
\caption{Mechanisms for nematic ordering: (a)  thermodynamic  (b) active.}
\label{fig:isonem}
\end{figure}

In a passive, equilibrium liquid crystal, the degree of orientational order is directly related to the inter-particle interactions and the temperature 
of the system. As the temperature is lowered the liquid crystal becomes more orientationally ordered to increase its positional entropy\cite{Frenkel2014}. This dependence of the order on temperature can be modelled phenomenologically through the material coefficients, $A$, $B$ and $C$ in the Landau-de Gennes bulk free energy \cite{P1995} (eqn~\ref{eqn:LDG}). In the absence of non-equilibrium effects, i.e. zero activity, the parameter $\overline{T} = 27A C / B^2$ controls the state of the system in the isotropic-nematic phase-space as shown schematically in Fig.~\ref{fig:isonem}(a). 

In an active nematic, the activity itself can provide a second route to nematic ordering, even in the absence of any thermodynamic tendency towards order i.e. at infinite temperature \cite{Thampi2015}(see Fig.~\ref{fig:isonem}(b)). This occurs because small fluctuations in the director field can lead to local nematic alignment which generates a net stresslet flow. For an extensile system, and $\lambda>0$, the flows can rotate and orient the nematogens, further strengthening the order in the system.

This mechanism is apparent from the continuum model of nematic liquid crystals when the momentum balance is primarily between active and viscous stresses, $2\mu E_{ij} = -\zeta Q_{ij}$. Then eqn.~\ref{eqn:evolution} may be rewritten as
\begin{equation}
\left(\partial_t+u_k\partial_k\right)Q_{ij}-\Omega_{ik}Q_{kj}+Q_{ik}\Omega_{kj} = \Gamma H^{\textnormal{active}}_{ij}
\label{activeorder}
\end{equation}
where
\begin{equation}
\begin{split}
H^{\textnormal{active}}_{ij} = K \nabla^2Q_{ij}+\frac{\lambda \zeta}{3\mu}Q_{ij}+\frac{\lambda \zeta}{\mu}\left[Q_{ik}Q_{kj}-Q_{pq} Q_{qp}\frac{\delta_{ij}}{3}\right] \\ -\frac{\lambda \zeta}{\mu}Q_{ij}Q_{pq}Q_{qp} 
\end{split}
\label{eqn:intrinsic}
\end{equation}
for $A = B = C = 0$. Terms in $\mathbf{H}^{\textnormal{active}}$ on the right hand side of eqn.~(\ref{activeorder}), which arise from the generalized advection term
$\mathbf{S}$, resemble those that appear in the molecular field $\mathbf{H}$ due to the Landau-de Gennes free energy $F_{\textnormal{LDG}}$ (see eqns.~(\ref{eqn:LDG}) and (\ref{eqn:potential})). Thus, activity acts as an effective free energy with the parameter $\frac{\lambda \zeta}{3 \Gamma \mu} > 0$ determining the strength of the ordering. %This transition is illustrated in Figure~\ref{fig:isonem}(b).

\begin{figure*}
\centering
\rotatebox{90} {\fbox{a) Unidirectional}}
\subfloat{}{\includegraphics[width=0.3\linewidth,trim={0 -2.5cm 0 0},clip]{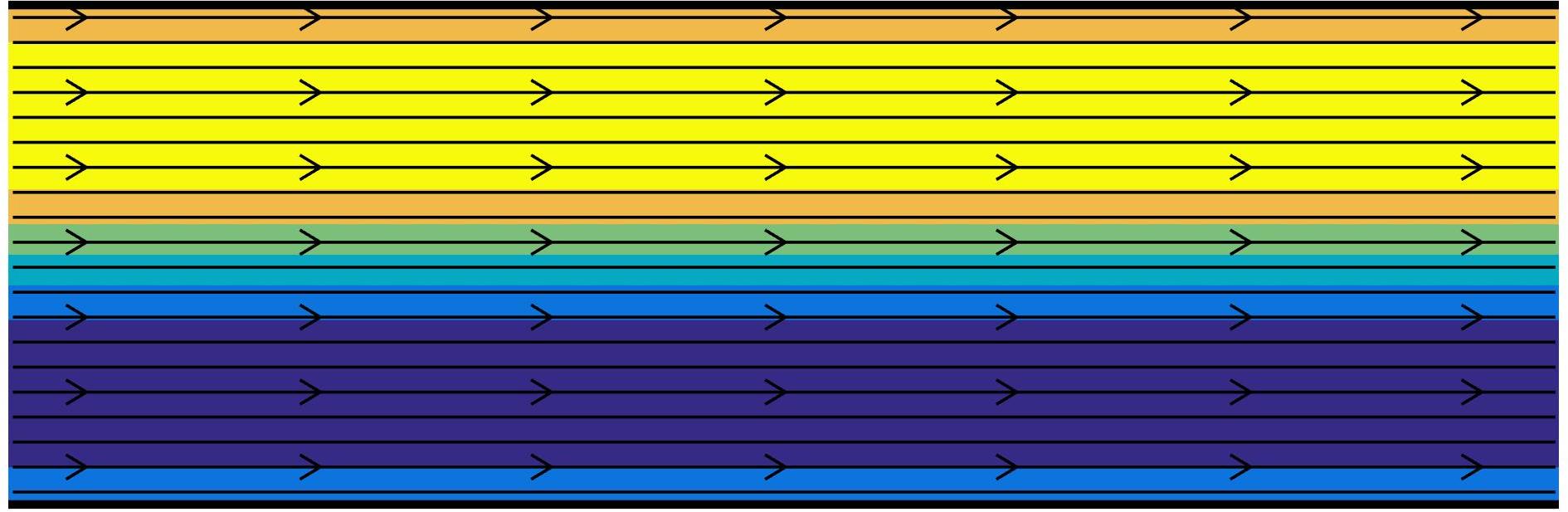}}\hspace{0.6cm}
\subfloat{}{\includegraphics[width=0.26\linewidth,trim={0 -1.5cm 0 0},clip]{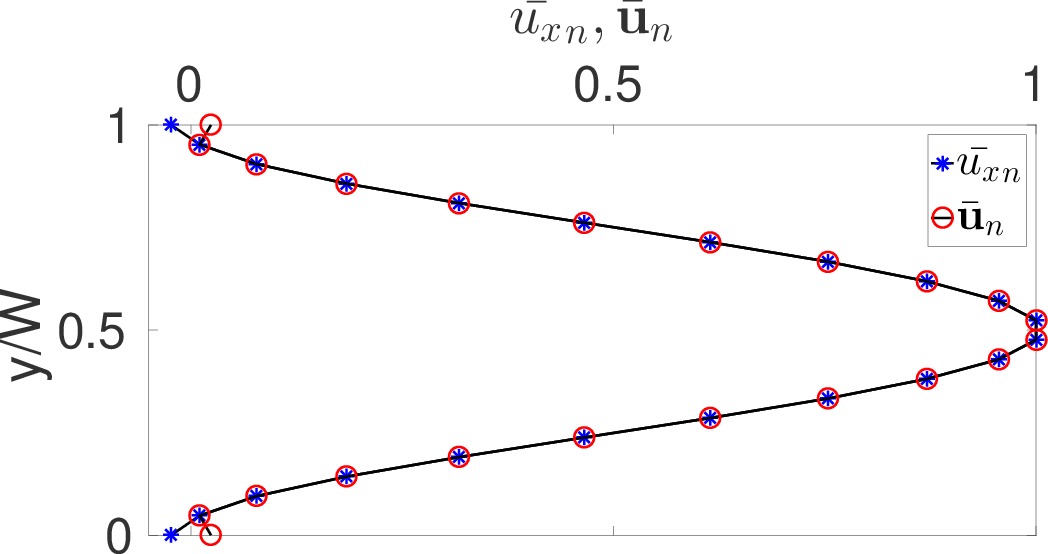}}\hspace{0.6cm}
\subfloat{}{\includegraphics[width=0.3\linewidth,trim={0 -2.5cm 0 0},clip]{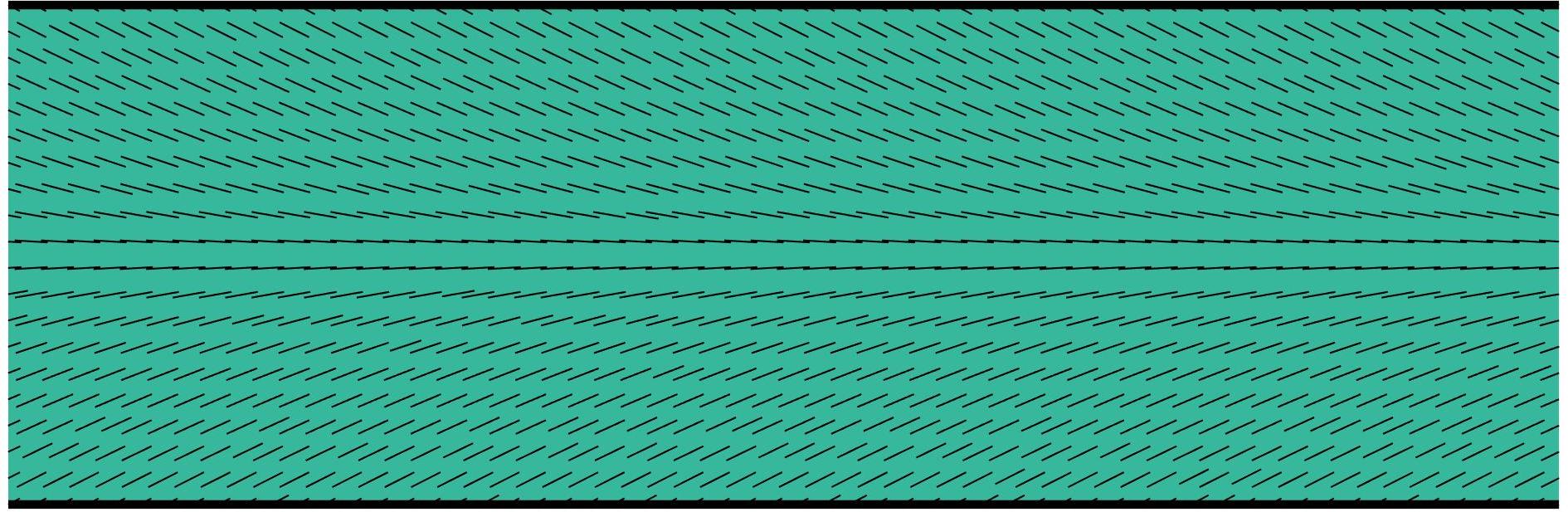}}
\vspace{0.1cm}
\\ \rotatebox{90} {\fbox{b) Oscillatory}}
\subfloat{}{\includegraphics[width=0.3\linewidth,trim={0 -1.5cm 0 0},clip]{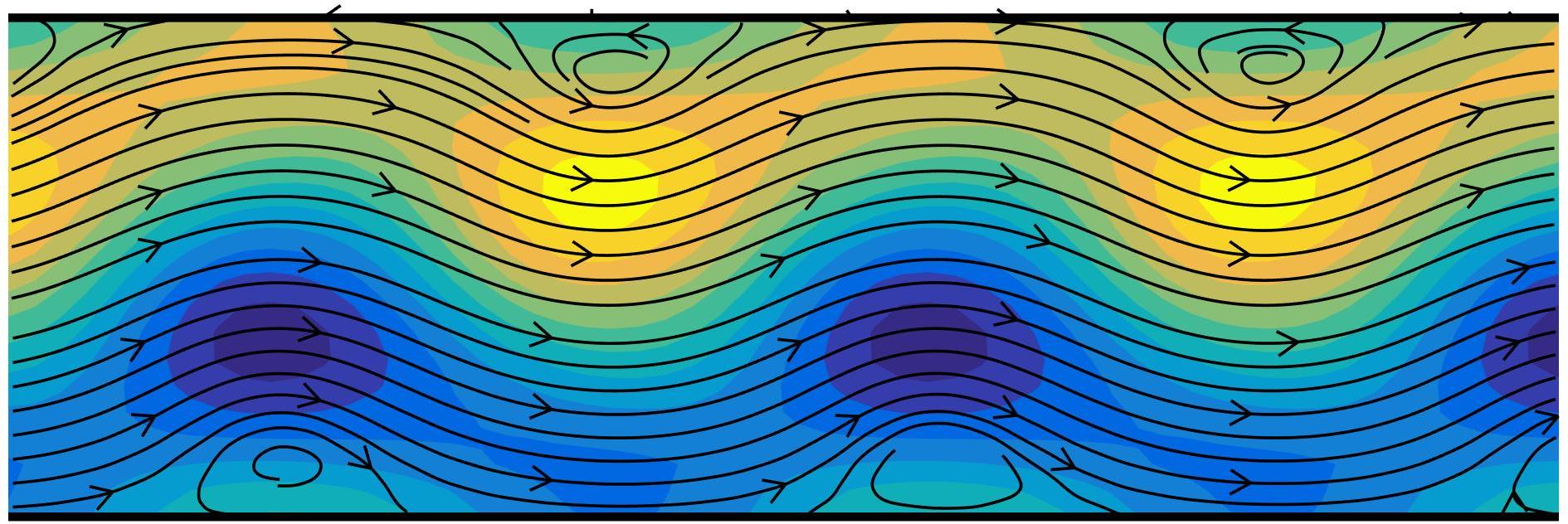}}\hspace{0.6cm}
\subfloat{}{\includegraphics[width=0.26\linewidth,trim={0 -1.5cm 0 1.55cm},clip]{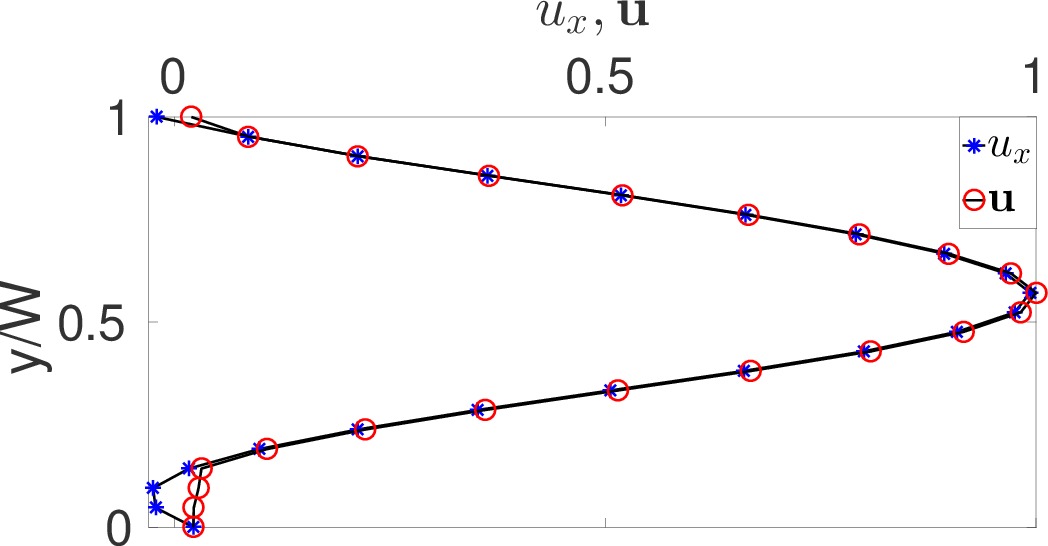}}\hspace{0.6cm}
\subfloat{}{\includegraphics[width=0.3\linewidth,trim={0 -1.5cm 0 0},clip]{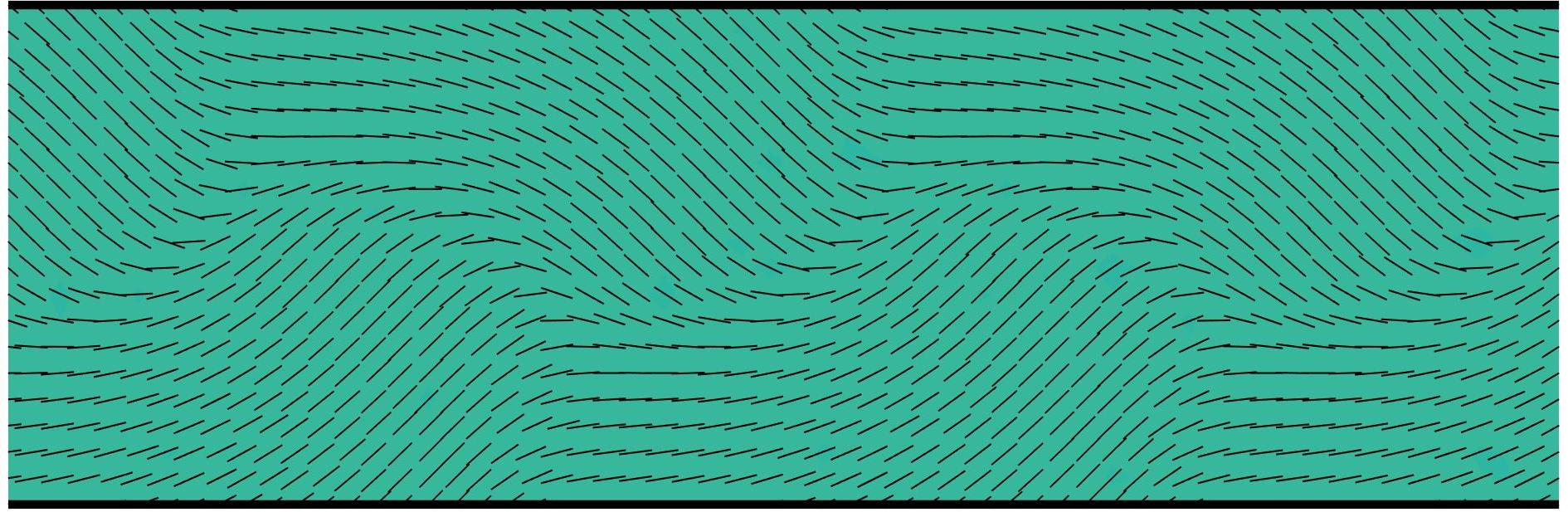}}
\vspace{0.1cm}
\\ \rotatebox{90} {\fbox{c) Dancing}}
\subfloat{}{\includegraphics[width=0.3\linewidth,trim={0 -0.5cm 0 0},clip]{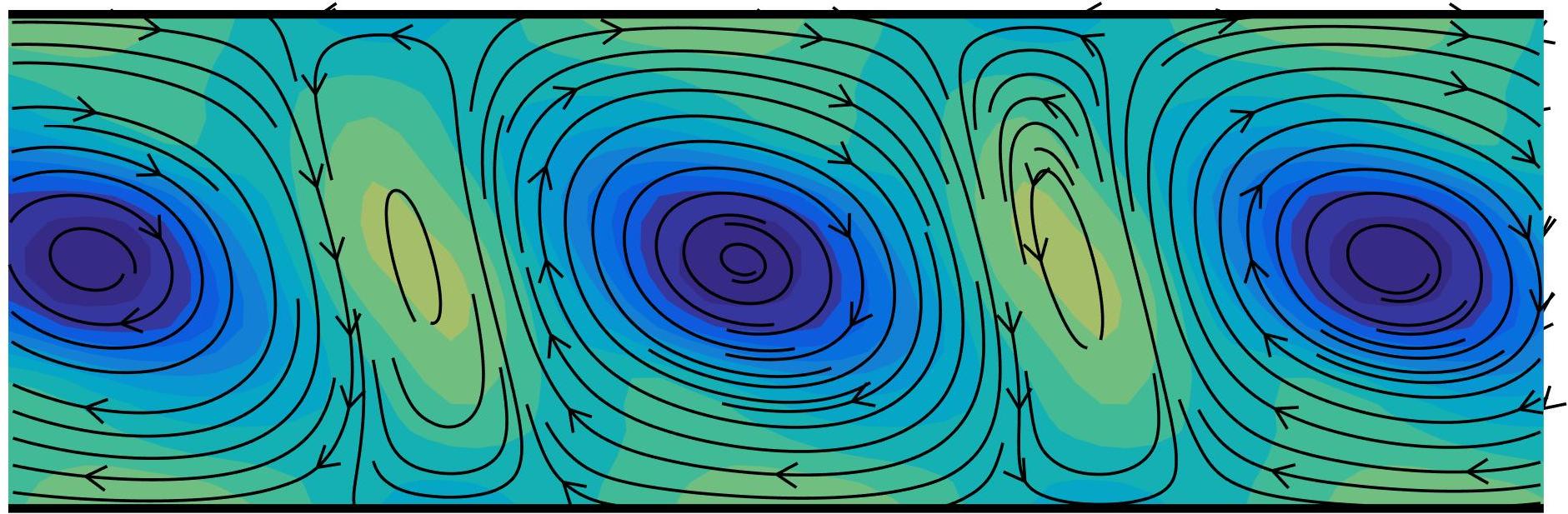}}\hspace{0.6cm}
\subfloat{}{\includegraphics[width=0.26\linewidth,trim={0 0 0 1.55cm},clip]{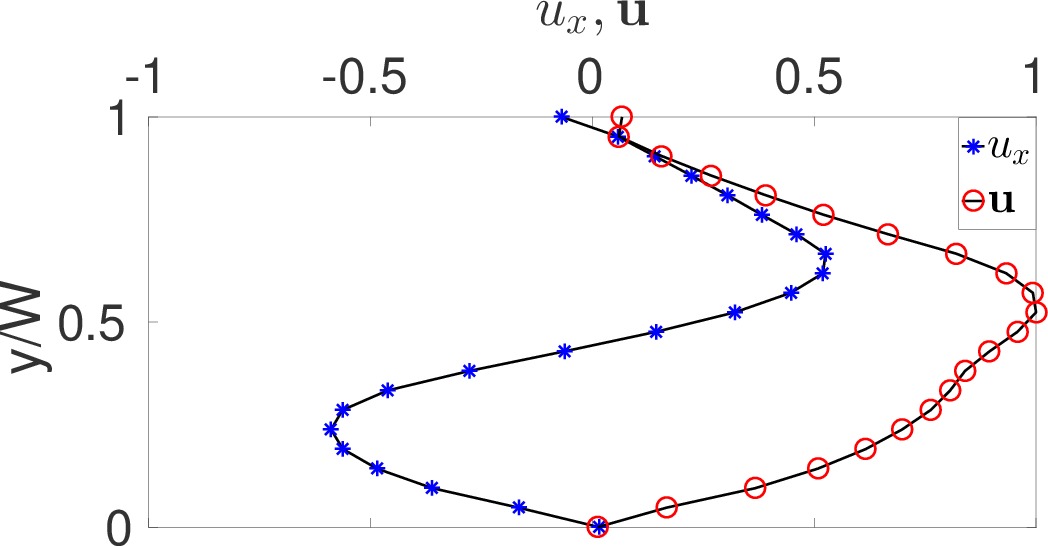}}\hspace{0.6cm}
\subfloat{}{\includegraphics[width=0.3\linewidth,trim={0 -0.5cm 0 0},clip]{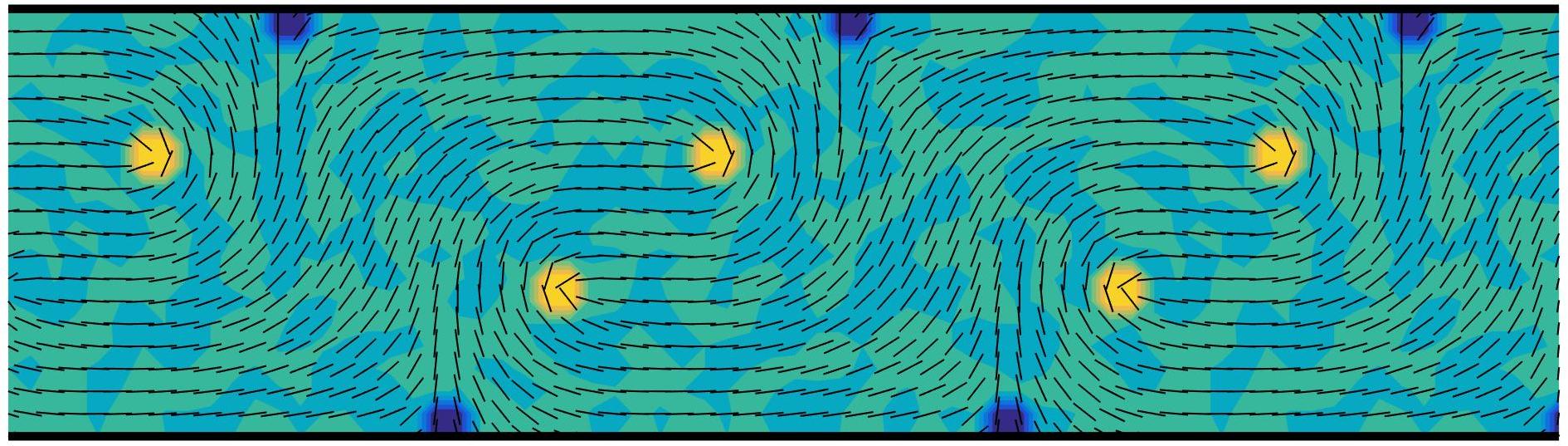}}
\vspace{0.1cm}
\\ \rotatebox{90} {\fbox{d) Turbulent}}
\subfloat{}{\includegraphics[width=0.3\linewidth,trim={0 -0.5cm 0 0},clip]{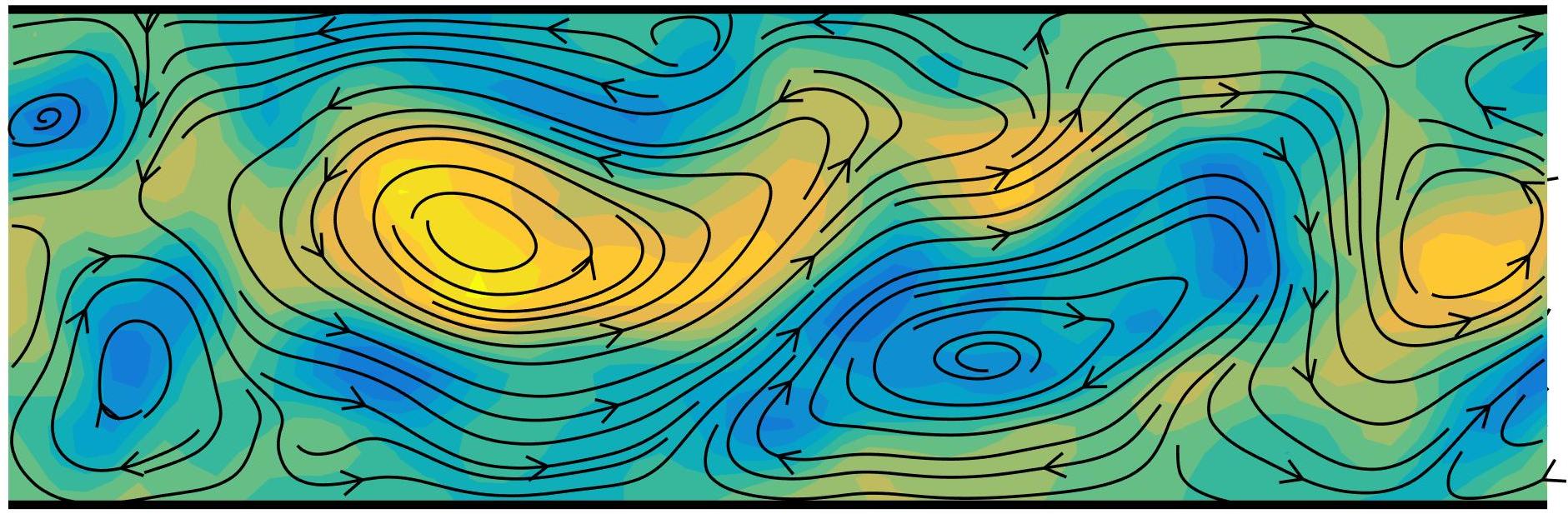}}\hspace{0.62cm}
\subfloat{}{\includegraphics[width=0.26\linewidth,trim={0 0 0 1.55cm},clip]{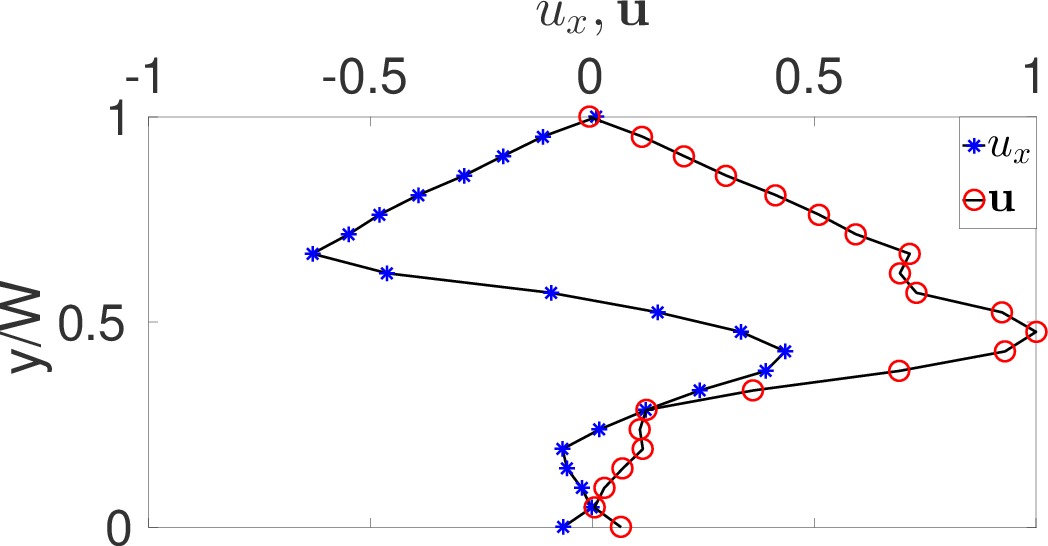}}\hspace{0.6cm}
\subfloat{}{\includegraphics[width=0.3\linewidth,trim={0 -0.5cm 0 0},clip]{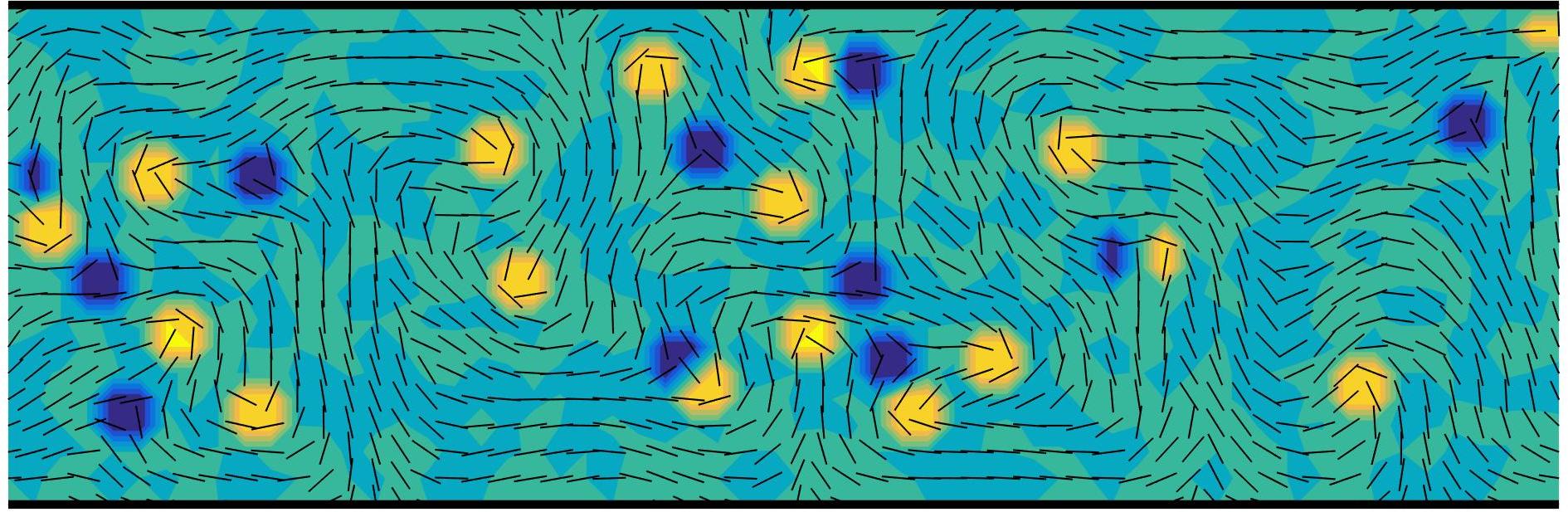}}
\vspace{0.5cm} \\ (i) \hspace{6cm} (ii) \hspace{5.5cm} (iii)
\caption{The four different flow states observed in a confined active nematic fluid. (a) unidirectional flow  at activity number = 9.79, (b) oscillatory flow at activity number = 10.95, (c) dancing flow at activity number = 18.97, (d) active turbulence at activity number = 40. The channel size is $100 \times 20$, the free energy parameters are $A=B=C=0$ (no thermodynamic ordering) and the flow aligning parameter is $\lambda=0.4$. For each flow configuration we show (i) the instantaneous flow streamlines and vorticity, normalised to their maximum value, with the colour scale denoting the variation from -1 (blue, clockwise) to $+1$ (yellow, anticlockwise). (ii) Instantaneous velocity profiles
${u}_x$ and ${|\textbf{u}|}$ at x = L/2 but normalized with maximum value of ${|\textbf{u}|}$. (iii) The director field showing $+1/2$ (yellow) and $-1/2$ (blue) defects. The same flow states occur in thermodynamically-stabilised systems.}
\label{fig:vel-p}
\end{figure*}

% \begin{figure}
% 	\centering
%      \subfloat[]{\includegraphics[width=\linewidth]{danc_new.eps}} \\
%     \subfloat[]{\includegraphics[width=\linewidth]{turb_new.eps}}
%         \caption{velocity profiles of dancing and turbulent states for 1 realisation but over 500 time steps.}
% \end{figure}

% \begin{figure}
% 	\centering
%      \subfloat[]{\includegraphics[width=\linewidth]{danc_def_new.eps}} \\
%     \subfloat[]{\includegraphics[width=\linewidth]{turb_def_new.eps}}
%         \caption{defect profiles for 1 realisation and over 500 timesteps.}
% \end{figure}

% \textcolor{LimeGreen}{(i) \textbf{section 4.2, to be confirmed case:-} At higher temperatures between activities 0.0085 and 0.009, there was no flow. Till activity = 0.0089 there was no flow and at 0.009 dancing state has occurred and at 0.0089 activity there are spurious velocities in the entire channel but the streamlines are in favour of forming a dancing state rather than an unidirectional flow state.\\
% (ii) Averaging over long times helped us in velocity distribution of dancing state but for turbulent state still it did not reach zero. \\
% (iii) Defect profiles did not change much as mentioned in fig.5. \\
% (iv) Running a metastable case in a higher length channel reduced the metastabilty.
% }

In general, both thermodynamic and flow ordering will contribute to determining the strength of the nematic order in the active system. In the following when we need to distinguish between the two cases we will use the terminology thermodynamically-stabilised active nematic to refer to $\overline{T} < \frac{27}{24}$ where the nematic state is (meta)stable in the passive system, and flow-stabilised active nematic to refer to $\overline{T} > \frac{27}{24}$ where activity-induced flow is needed to stabilise the nematic order.

\section{Active flows  in a channel}
\label{sec:fourstates}

\subsection{Flow states} \label{sec:flow_states}

We next describe the different flow configurations that can be generated when an  active nematic is confined within a long channel.  
For small activity there is no motion in a channel of finite width; the fluid undergoes a spontaneous flow transition at a threshold activity $\zeta_c$\cite{Voituriez2005}. Depending upon parameters, four distinct types of flow fields can be obtained as activity is increased beyond this threshold. These states, (i) unidirectional (ii) oscillatory (iii) `dancing' flow and (iv) turbulent flows, are shown in Fig.~\ref{fig:vel-p}. Change in activity is reported in terms of the activity number, $W\sqrt{{\zeta}/{K}}$ which is the ratio of the two characteristic length scales, the channel width and the length scale associated with active vortices, $\sqrt{{K}/{\zeta}}$.

\noindent
{\bf Unidirectional flow,} illustrated in Fig.~\ref{fig:vel-p}(a) for an activity slightly larger than $\zeta_c$, is characterised by streamlines aligned parallel to the channel walls. The velocity field is translationally invariant along $x$ and ${u}_x = {|\mathbf{u}|}$ varies from zero at the channel walls to a maximum in the centre of the channel. Topological defects are not generated in unidirectional flows.

\noindent
{\bf Oscillatory flow:} As the activity is increased the component of velocity normal to the channel walls becomes non-zero. However, $u_y$ is typically smaller than $u_x$. The corresponding streamlines, vorticity field and director field are shown in Fig.~\ref{fig:vel-p}(b). An alternating pattern of vorticity patches with opposite sign starts to develop, and the corresponding director field shows locally nematic regions. 
%Since oscillatory flows have a nonzero $u_y$, both $\bar{u}_x$ and $|\mathbf{u}|$  are only approximately parabolic. Profile for $\bar{q}$ is similar to that of unidirectional flows and 
Topological defects are occasionally present.

\begin{figure}
	\centering
     \subfloat[]{\includegraphics[width=\linewidth]{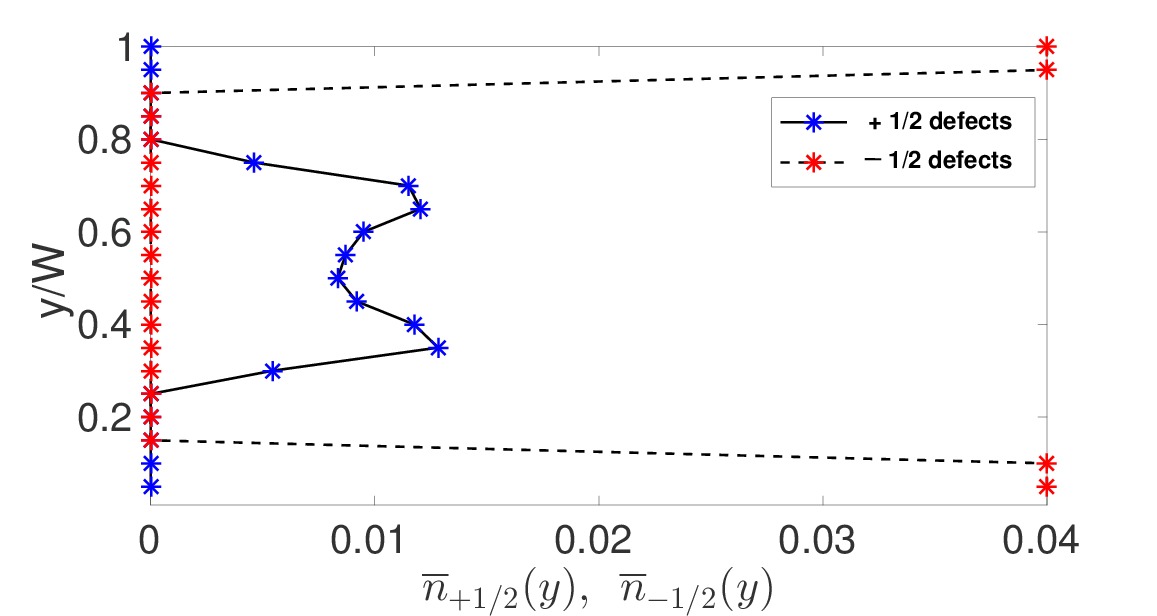}} \\
    \subfloat[]{\includegraphics[width=\linewidth]{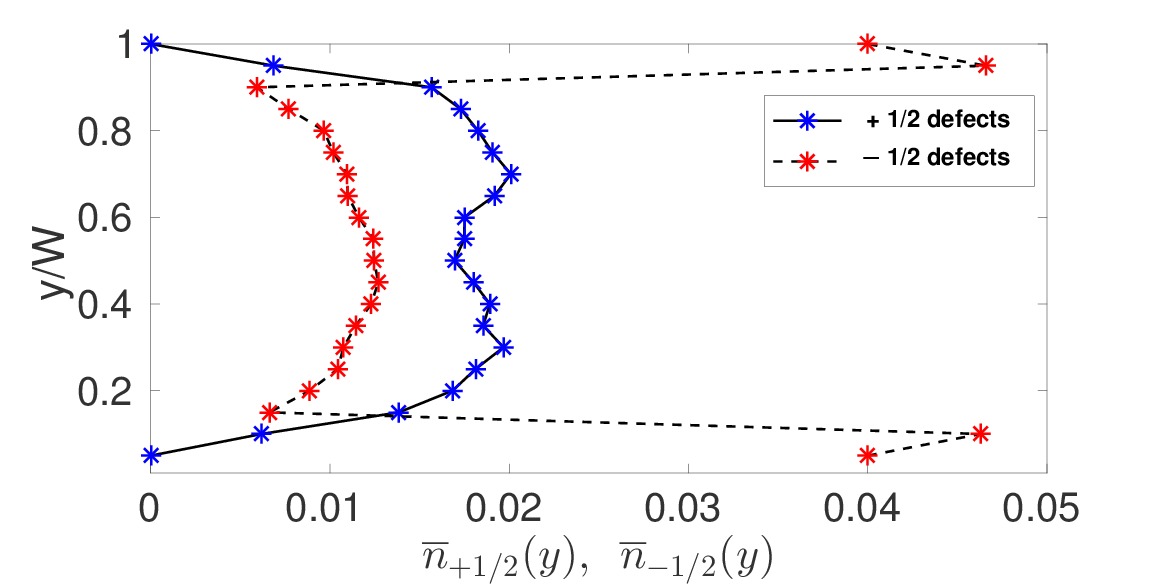}}
        \caption{Distribution of $\pm 1/2$ defects across the channel in (a) the dancing state and (b) the turbulent state corresponding to Fig. 3(c) and Fig. 3(d) respectively. $\bar{n}_{\pm 1/2}(y) = 
\langle  \delta(x-x_{\pm 1/2},y) \rangle$ where  $\langle \cdots \rangle$ denotes an average over the channel length, time and 10 realizations, $x_{\pm 1/2}$ is the $x$-location at which $\pm 1/2$ defects are present and $\delta$ is the discrete delta function.}
    \label{fig:defects}
\end{figure}

\noindent
{\bf Dancing flow:}\cite{Shendruk2017} A one-dimensional flow vortex lattice, with topological defects following regular oscillatory paths along the channel, is observed at higher activities when the length scale of active vortices, which decreases with increasing activity, becomes of order the channel width (Fig.~\ref{fig:vel-p}(c)). The channel is filled by a row of velocity vortices which alternate in direction.
In contrast to the unidirectional and oscillatory states, here the director field develops a regular pattern of motile topological defects. 
%Once the steady state is established there is no further creation or annihilation of the defects and the system is stabilised into a dynamically ordered configuration. 
$-\frac{1}{2}$ defects are confined close to the walls due to elastic interactions with the boundaries. However, because of their innate self-propulsion, $+\frac{1}{2}$ defects continuously move along the channel, in either direction, along well-defined, oscillatory trajectories. %The movement of the $+{1}/{2}$ defects is very similar to partner exchange in Ceilidh dance and hence following \cite{shendruk2017dancing} we will call it dancing state. 
Fig.~\ref{fig:defects}(a) compares the distribution of $+\frac{1}{2}$ and $-\frac{1}{2}$ defects across the channel. Once the oscillation is established there is no nucleation/annihilation of topological defects. As the defects move there is an accompanying oscillation in the relative strengths of the positive and negative vortices. Hence ${u}_x$, and to a lesser extent ${|\mathbf{u}|}$ vary with time.\\
\noindent
{\bf Turbulent flows}, with properties summarised in Fig.~\ref{fig:vel-p}(d), are seen when the typical size of the active vortices is smaller than the channel width. 
Flow vortices, which are not uniform in size \cite{Guillamat2017,Giomi2015}, are short-lived and move around in the channel. The resulting streamlines appear very similar to those of high Reynolds number turbulent flows, hence the name active or mesoscale turbulence \cite{Wensink2012}. $\pm\frac{1}{2}$ defects are continuously created in the bulk of the fluid, move around and then undergo annihilation. The defect distribution, illustrated in Fig.~\ref{fig:defects}(b), shows less asymmetry in the locations of $\pm\frac{1}{2}$ defects, although the -1/2 defects still tend to reside near the walls.
%In the turbulent state $\bar{u}_x$ averages to zero over time, but $\overline{|\mathbf{u}|}$ is non-zero, and has a maximum at the centre line. 

\subsection{Flow order parameters}

\begin{figure}
	\centering
		\includegraphics[width=1.1\columnwidth,trim={2cm 0 2cm 0},clip]{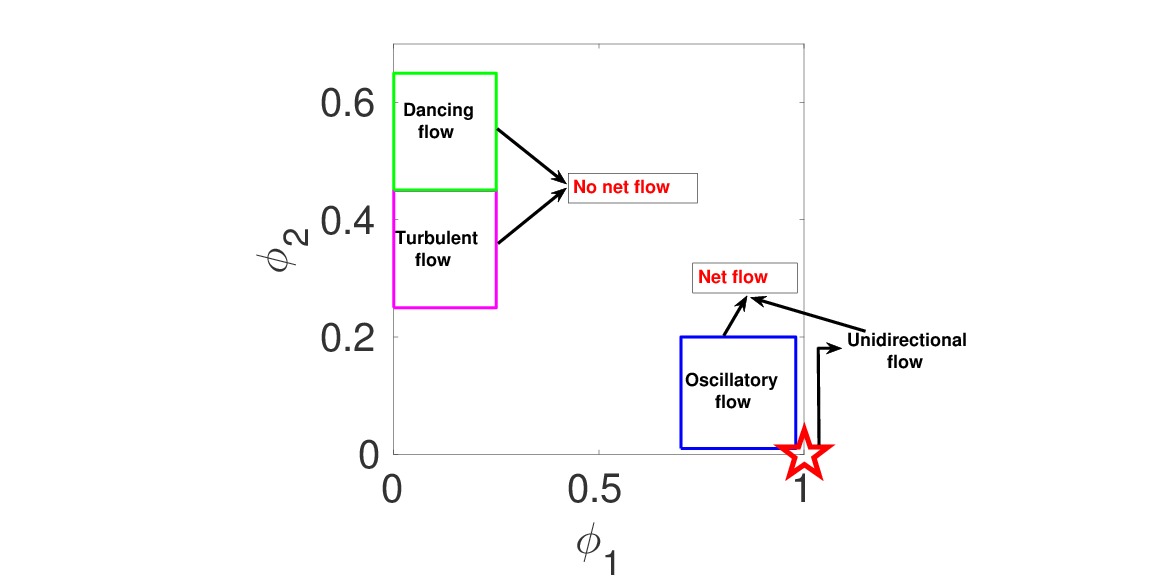} 
		\caption{Identification of the different flow states based on the values of the flow order parameters $\phi_1$ and $\phi_2$ (eqns.~(\ref{eqn:phi1}) and (\ref{eqn:phi2})).}
		\label{fig:floworder}
\end{figure}

In the previous section we have seen that field variables are different for the different flow states. Therefore, to distinguish between the different flow configurations in the simulation, it is helpful to define order parameters that estimate the relative net flow along $x$ and along $y$
\begin{align} \label{eqn:phi1}
\phi_1= \left\langle\left|\left\langle \frac{u_x(x,y)}{|\mathbf{u}(x,y)|} \right \rangle _{x}\right|\right\rangle_y,
\end{align}
\begin{align} \label{eqn:phi2}
\phi_2= \left\langle\left|\left \langle \frac{u_y(x,y)}{|\mathbf{u}(x,y)|} \right \rangle _{y}\right|\right\rangle_x.
\end{align}
where $\langle \cdots \rangle_{x}$, $\langle \cdots \rangle_{y}$ denote averages along the length and width of the channel.
For unidirectional flows, $u_y = 0$ everywhere and thus $\phi_1 = 1, \phi_2 = 0$. In case of oscillatory flows, $u_x$ varies throughout the channel, but it dominates over $u_y$. We identify $0.7 < \phi_1 < 1$ and $\phi_2<0.2$ as oscillatory flows in the simulations. 
For the dancing state of alternating vortices neither $u_x$ nor $u_y$ is negligible. However, $u_y$ is persistent across the channel and $u_x$ changes direction along the channel. Visual inspection shows that small $\phi_1$ and $0.45 < \phi_2 < 0.65$ identifies the dancing states. For turbulent flow, both $u_x$ and $u_y$ fluctuate throughout the channel yielding $\phi_1$ small and $\phi_2 < 0.45$. These identifications are summarised in  Fig.~\ref{fig:floworder}.

\section{Results} \label{sec:results}

We now present results showing how the flow configurations change as the activity number and the aligning parameter are varied. The aligning parameter $\lambda\propto(r^2-1)/(r^2+1)$ is related to the aspect ratio of the active particles, $r$. We consider 
$\lambda>0$, corresponding to  rods.
Results for thermodynamically-stabilised active nematics will be presented first, followed by a comparison to those for flow-stabilised fluids. We will then 
discuss the relationship between the two cases.

The flow configurations identified at each point in the phase plot are obtained from $10$ different realisations of each simulation. Different realisations were achieved by changing the seed that prescribes the random distribution of the director field in the initial conditions. We found that  the final flow state is often strongly dependent on the initial conditions, and we indicate this by using two or more superimposed symbols. (This metastability was less pronounced in reference \cite{Shendruk2017} because of the nematic initial conditions of the director field.)
Moreover unidirectional flow states only exist over a very narrow range of parameters. Therefore, for clarity, they are not distinguished from oscillatory flow states in the phase plots. 

\subsection{Flow states of a thermodynamically-stabilised active nematic}
\label{sec:lower_T}

\begin{figure} 
	\centering
    \includegraphics[width=\columnwidth,trim={2.15cm 0 3cm 0},clip]{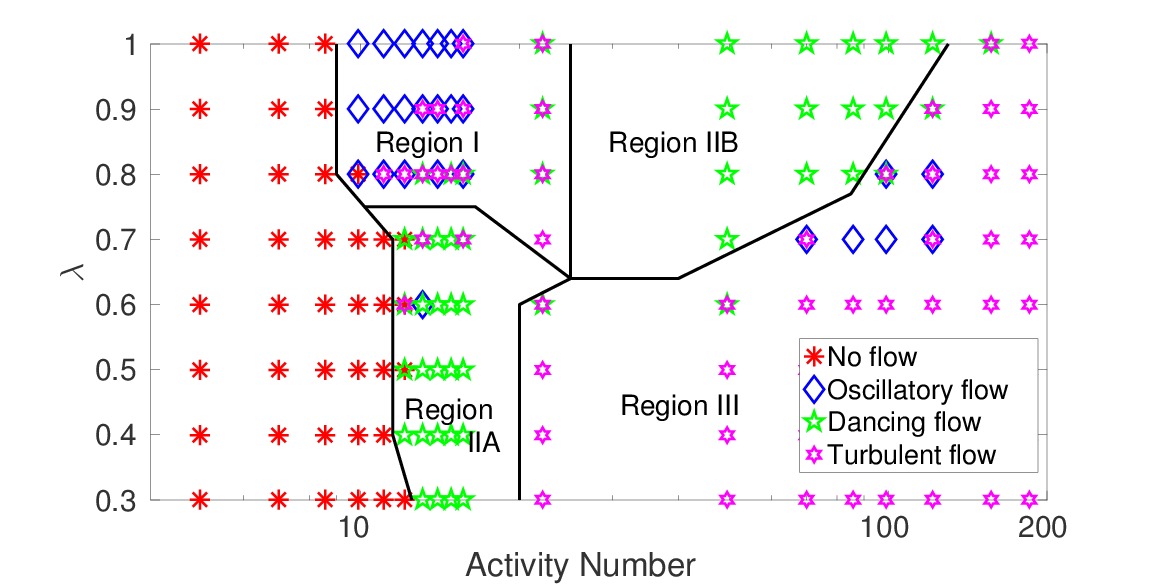}\label{fig:ametastable}
    \caption{Flow states of a confined, thermodynamically-stabilised, active nematic $( A=0.005, B = C = 0.3$ which corresponds to $\overline{T} = 27/60$) as a function of activity number, $W\sqrt{{\zeta}/{K}}$, and aligning parameter, $\lambda$. Black lines are  a guide to eye to demarcate regions corresponding to different flow states. } 
    %(b) at various temperatures but at a given $\lambda = 0.7$.}
    \label{fig:metastable}
\end{figure}

Fig.~\ref{fig:metastable} shows the different flow states of a thermodynamically-stabilised active nematic fluid as a function of activity number and aligning parameter. 
For sufficiently high aligning parameter ($\lambda > \sim 0.7$), the phase sequence with increasing activity number is no flow $\rightarrow$ oscillatory flow $\rightarrow$ dancing flow $\rightarrow$ active turbulence. However, for smaller $\lambda < \sim 0.6$, the quiescent state transitions directly to the dancing state and then, at a higher activity number, to turbulence.

At many points in the phase diagram, and particularly close to the borders between different flow states, the final steady state depends sensitively on the initial conditions. Interestingly, at the borders, the flows obtained are not always the ones pertaining to the neighbouring regions: thus the smearing of the boundaries is not just a consequence of hysteresis associated with the transitions.  For example, turbulent flow occurs between the oscillatory (region I) and dancing (region IIB) flow states; and the oscillatory state, which corresponds to a net flow, occurs between the dancing (region IIB) and turbulent (region III) regimes, which do not generate any net flow.

\subsection{States of a flow-stabilised active nematic} 
\label{sec:higher_T}

\begin{figure}
	\centering
			\includegraphics[width=\linewidth,trim={2.15cm 0 3cm 0},clip]{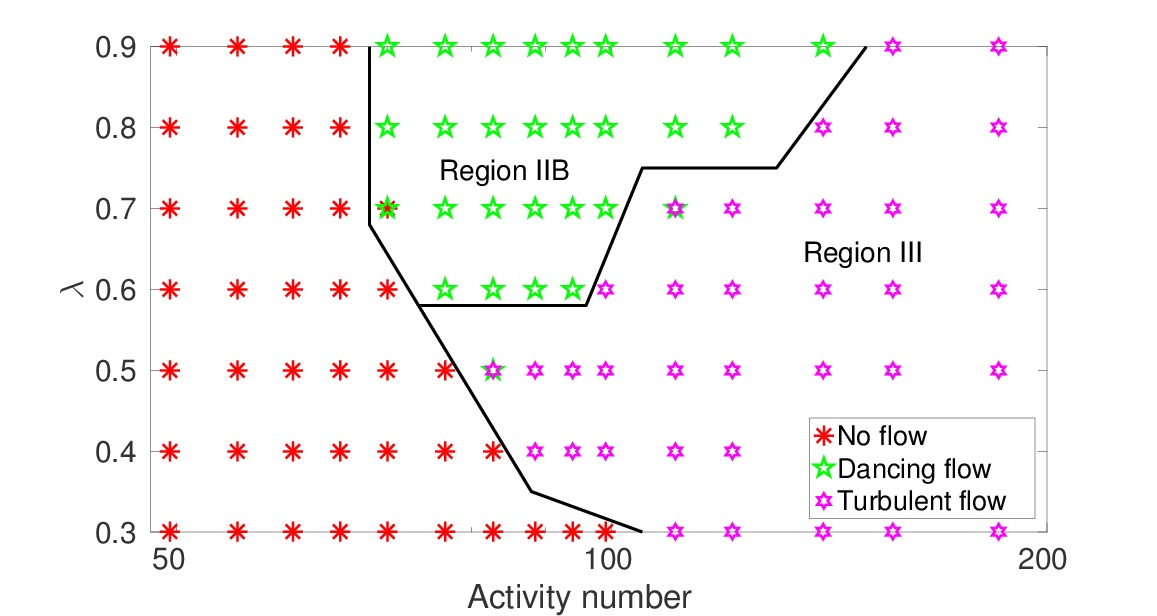}
		\caption{Flow states of a confined, flow-stabilised, active nematic $(A = 0.015, B = C = 0.3$ which corresponds to $\overline{T} = 27/20$) as a function of activity number, $W\sqrt{{\zeta}/{K}}$, and aligning parameter, $\lambda$. Black lines are  a guide to eye to demarcate regions corresponding to different flow states.}
		\label{fig:flow_higher_T}
\end{figure} 

We now compare results for an active nematic where the ordering is predominantly from the extensile flow field ($\overline{T}=\frac{27}{20}$). A phase plot showing the stable flow configurations as a function of activity number and aligning parameter is presented in Fig.~\ref{fig:flow_higher_T}. For $\lambda > \sim 0.7$ the phase sequence with increasing activity number is no flow $\rightarrow$ dancing flow $\rightarrow$ turbulence. For $\lambda < \sim 0.6$ the quiescent state transitions directly to turbulence.

There are several differences between the behaviour in the thermodynamically-aligned and flow-aligned nematics. In the latter case the stationary state (no flow state) becomes unstable at a considerably larger value of the activity number, as the activity has to drive first ordering, and then the instability to flow. Once flow is possible the dancing or turbulent states are more likely to be stabilised directly and no unidirectional or oscillatory flow states were observed within the range of parameters studied. %\sout{Although in Fig.~\ref{fig:flow_higher_T} there are no indications of oscillatory or unidirectional flow states, we must note that oscillatory flows were occasionally observed for the flow-stabilised region, but over a very narrow region of activity numbers and with a very small probability.} 
Another striking difference is that, unlike the thermodynamically-stabilised regime, the final flow states in the flow-stabilised case were independent of initial conditions for all the parameters we tested.

By varying the effective temperature, $\overline{T}$, it is possible to crossover from a system where the nematic ordering is predominantly thermodynamically driven  to one where it is predominantly due to the flow. Fig.~\ref{fig:bmetastable} shows how the flow states vary with temperature for different activity numbers for alignment parameter $\lambda=0.7$.  
The threshold activity number for flow generation is $\approx 10$ for $\overline{T} < \sim 1$. The threshold sharply increases around $\overline{T} = 1$ and then continues to increase mildly with further increase in temperature.
The oscillatory states become unstable for higher temperature, and metastability decreases with increasing temperature. However, the pattern of states is complex. Therefore, in the next subsection, we show that a different choice of variables allows us to present a more unified picture of the results for varying temperatures.

\begin{figure}
	\centering
 	 \includegraphics[width=1.1\columnwidth,trim={1.75cm 0 0 0},clip]{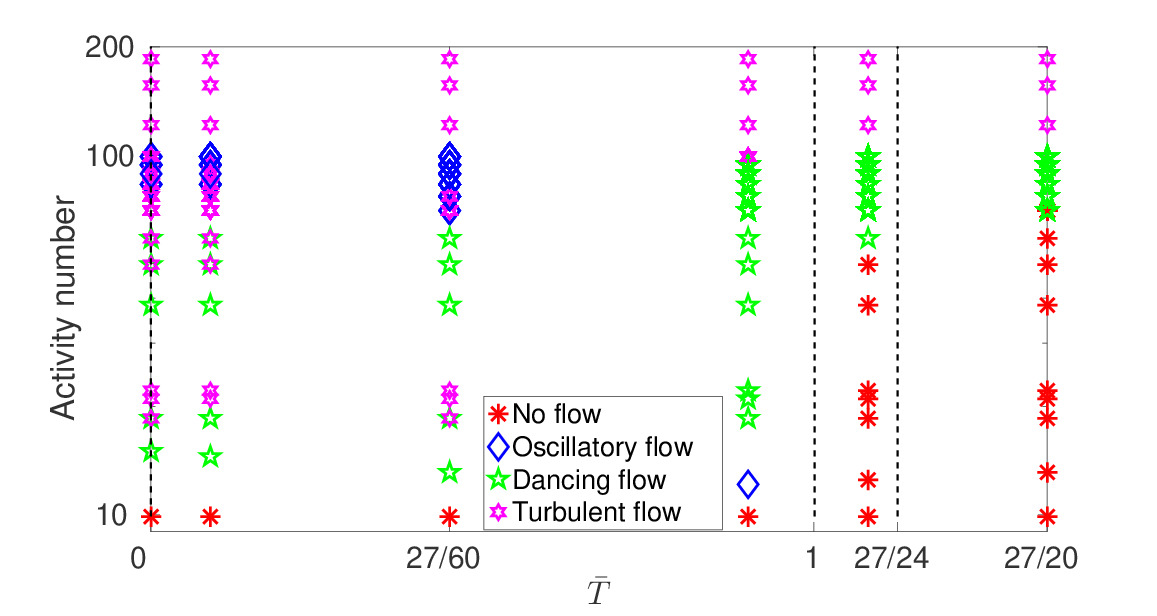}\label{fig:bmetastable}
     	\caption{Flow states of a confined, active nematic  as a function of effective temperature $\overline{T}$, and activity number, $W\sqrt{{\zeta}/{K}}$, for aligning parameter, $\lambda=0.7$. }
		\label{fig:bmetastable}
\end{figure}
    
\subsection{Unifying the observations at all temperatures} 

Motivated by the dependence of the flow configurations on both the $\lambda$ and the degree of nematic ordering that is induced by thermodynamics and flow, we recall that the response of a nematic rod to flow gradients depends on both  $\lambda$ and on the magnitude of the order parameter $q$.
If the order parameter is assumed to be of a constant magnitude then we may write~\cite{Marenduzzo2007}
\begin{align}
\lambda_1 = \left(\frac{3q+4}{9q}\right) \lambda .
\label{eqn:lambda1}
\end{align}
The director field aligns at a constant angle to a simple shear flow when $\lambda_1 > 1$, otherwise it tumbles\cite{Edwards2009}. We will now investigate whether $\lambda_1$ provides a more suitable control parameter for the flow behaviour.

To obtain a representative value of the magnitude of order parameter for each simulation, we define $q_{avg} = \langle \langle q \rangle_x \rangle_y $, averaging over the simulation domain and time.  No-flow configurations are excluded from this calculation as $q_{avg} = 0$ for these states for $\overline{T}>1$. 

\begin{figure}
	\centering
			\includegraphics[width=1.1\linewidth,trim={1.5cm 0 0 0},clip]{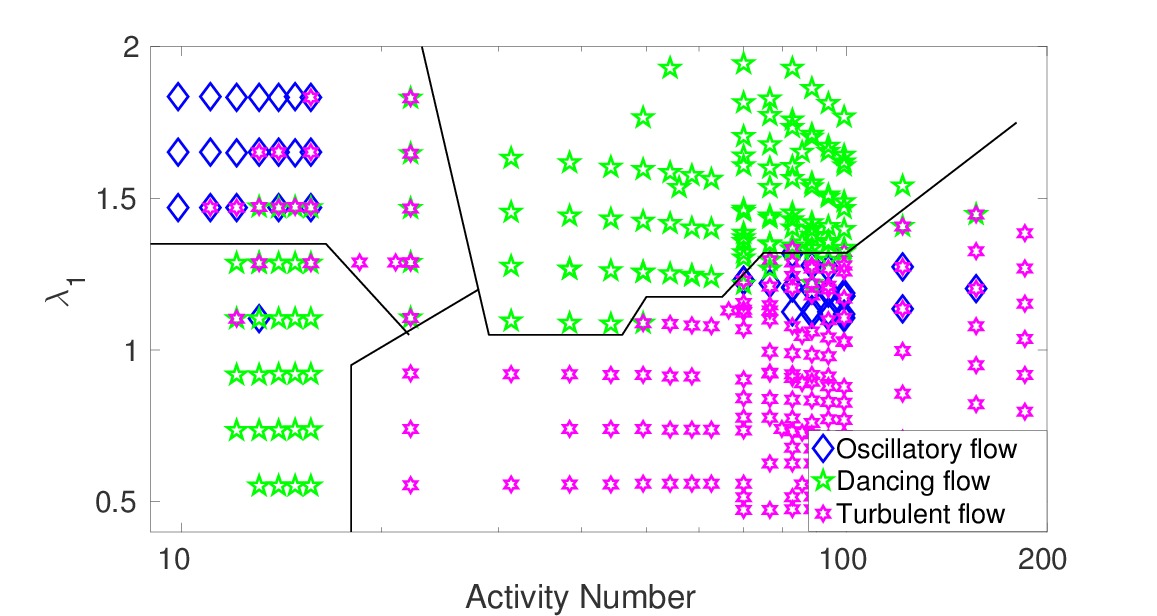}
		\caption{Flow states of a confined active nematic  as a function of activity number, $W\sqrt{{\zeta}/{K}}$, and scaled aligning parameter, $\lambda_1$, including data from all temperatures. Black lines are  a guide to eye to demarcate regions corresponding to different flow states.}
		\label{fig:L1_vs_An}
\end{figure}

A phase plot showing the stable flow configurations as a function of activity number and $\lambda_1$ is presented in Fig.~\ref{fig:L1_vs_An}. This figure includes data from Figs.~\ref{fig:metastable} - \ref{fig:bmetastable}. Note that the results from higher temperatures lie on the right-hand side of the plot because spontaneous flow only occurs for larger activity numbers.

To a good approximation the flow states divide into four regions. For flow aligning systems, $\lambda_1 > \sim 1$, there is a transition from oscillatory to dancing flow with increasing activity number, whereas for flow tumbling nematics, $\lambda_1 < \sim 1$, the transition is from dancing to turbulent flow. This is reasonable as a tumbling configuration will more easily form flow vortices.

 Thus we are able to  present  the results for thermodynamically-stabilised and flow-stabilised active nematics within the same framework. Our results show that the key parameters controlling the transitions are the activity number, and whether the nematogens are flow-aligning or flow-tumbling. The origin of the nematic ordering (thermodynamic- or flow-stabilisation) is not important.

\section{Summary and discussion}

Typically flow instabilities in active nematics lead to a chaotic flow field characterised by high vorticity and motile topological defects. In the previous work a flow lattice was observed either in a flow stabilized active nematic laid on a substrate \cite{Doostmohammadi2016} or in a thermodynamically stabilized, tumbling active nematic confined between two plates \cite{Shendruk2017}. Here we show that for both thermodynamically and flow stabilised active nematics regular flows can be obtained by confinement. In two dimensions these include laminar flow, oscillatory flow and a `dancing' flow state where +1/2 topological defects move along the channel. In this paper we have presented results showing the region of stability of the different flow configurations, concentrating in particular on comparing systems below and above the isotropic-nematic transition, $\overline{T}=1$, where the nematic ordering is predominantly due to thermodynamic forces or active flow stabilisation, respectively. This was motivated by noting that many experimental systems are not nematic until active forcing is introduced.
Similar flow states are seen at all temperatures, but there is a much lower probability of obtaining net flow for $\overline{T}>1$.
We found that the final flow configuration is strongly dependent on the choice of initial conditions for $\overline{T}<1$. However, this metastability is absent at higher temperatures. Moreover the critical activity required to trigger hydrodynamic instabilities is much higher above $\overline{T}=1$. This is because the activity has to first stabilise local nematic order before a spontaneous flow transition can occur. 
%Spontaneous hydrodynamic instabilities associated with an active nematic fluid has been well established in the literature. Here we show that similar flow states are also obtained for an active paranematic fluid. An active paranematic fluid does not have any inherent orientational order in the absence of activity. Activity generates nematic order and surprisingly, heterogeneities in the order to generate coherent fluid flows. In particular unidirectional, oscillatory, dancing and turbulent flow states are observed for an active paranematic fluid confined between two parallel plates.

By varying the temperature, activity and flow alignment we found that the key parameters controlling the transition are the activity number and a scaled alignment parameter.  The activity number, $W\sqrt{{\zeta}/{K}}$, is a dimensionless variable characterising the ratio of the channel width to the vortex length scale associated with the active fluid. Larger activity numbers correspond to smaller vortices and hence increases in the activity number favour first the dancing state and then active turbulence. The scaled alignment parameter, $\lambda_1$, takes into account both the shape of the nematogens and the magnitude of the nematic order and controls whether the system is flow aligning or flow tumbling. The dancing and turbulent states appear at lower activity numbers in flow tumbling systems because it is easier to form vortices.

The transition to laminar flow has been observed in confined confluent cell layers~\cite{Duclos2018}, and there have been indications of vortex structures in experiments studying both cells\cite{Nier2016,Duclos2018}and bacteria\cite{Wioland2016}. It would be of interest to look at other active systems in channels, although experiments may be difficult due to metastability effects. Recent experiments on microtubule gels in three-dimensional channels with rectangular cross sections have shown that the existence of net flow depends on the aspect ratio, but not the dimensions, of the channel cross section\cite{Wu2017}. This system is in the flow-aligning regime, above the isotropic-nematic transition temperature, and extending our simulations to three dimensions might help to explain this surprising result.

%Confinement has been projected as a way to control the active fluid flow in microfluidic applications \cite{theillard2017geometric}. Recent observations show how the aspect ratio of cross sectional area of a channel affects the active flows. Our simulations take a deeper investigation (though in two dimensions) of analyzing the role of both aspect ratio of active constituents (directly related to the tumbling parameter) and the existence of an intrinsic orientational order in determining the flow states. Thus, our simulations also complement the experimental investigations in finding the ways to control and tune the active fluid flows for microfluidics applications.

%\ju{Can we bill the zero free energy as infinite temperature? 
%Actually both mechanisms are always happening except at infinite T. The difference below $T_c$ is that the thermodynamic ordering is enough to give long range order}

%\ju{WHY $\lambda_1$ controls the collapse}

 %%%REFERENCES%%%
\bibliography{references3} %You need to replace "rsc" on this line with the name of your .bib file

\providecommand*{\mcitethebibliography}{\thebibliography}
\csname @ifundefined\endcsname{endmcitethebibliography}
{\let\endmcitethebibliography\endthebibliography}{}
\begin{mcitethebibliography}{44}
\providecommand*{\natexlab}[1]{#1}
\providecommand*{\mciteSetBstSublistMode}[1]{}
\providecommand*{\mciteSetBstMaxWidthForm}[2]{}
\providecommand*{\mciteBstWouldAddEndPuncttrue}
  {\def\EndOfBibitem{\unskip.}}
\providecommand*{\mciteBstWouldAddEndPunctfalse}
  {\let\EndOfBibitem\relax}
\providecommand*{\mciteSetBstMidEndSepPunct}[3]{}
\providecommand*{\mciteSetBstSublistLabelBeginEnd}[3]{}
\providecommand*{\EndOfBibitem}{}
\mciteSetBstSublistMode{f}
\mciteSetBstMaxWidthForm{subitem}
{(\emph{\alph{mcitesubitemcount}})}
\mciteSetBstSublistLabelBeginEnd{\mcitemaxwidthsubitemform\space}
{\relax}{\relax}

\bibitem[Kruse \emph{et~al.}(2004)Kruse, Joanny, J{\"u}licher, Prost, and
  Sekimoto]{Kruse2004}
K.~Kruse, J.~F. Joanny, F.~J{\"u}licher, J.~Prost and K.~Sekimoto,
  \emph{{P}hys. {R}ev. {L}ett.}, 2004, \textbf{92}, 078101\relax
\mciteBstWouldAddEndPuncttrue
\mciteSetBstMidEndSepPunct{\mcitedefaultmidpunct}
{\mcitedefaultendpunct}{\mcitedefaultseppunct}\relax
\EndOfBibitem
\bibitem[Schaller \emph{et~al.}(2010)Schaller, Weber, Semmrich, Frey, and
  Bausch]{Schaller2010}
V.~Schaller, C.~Weber, C.~Semmrich, E.~Frey and A.~R. Bausch, \emph{{N}ature},
  2010, \textbf{467}, 73--77\relax
\mciteBstWouldAddEndPuncttrue
\mciteSetBstMidEndSepPunct{\mcitedefaultmidpunct}
{\mcitedefaultendpunct}{\mcitedefaultseppunct}\relax
\EndOfBibitem
\bibitem[Saw \emph{et~al.}(2017)Saw, Doostmohammadi, Nier, Kocgozlu, Thampi,
  Toyama, Marcq, Lim, Yeomans, and Ladoux]{Saw2017}
T.~B. Saw, A.~Doostmohammadi, V.~Nier, L.~Kocgozlu, S.~P. Thampi, Y.~Toyama,
  P.~Marcq, C.~T. Lim, J.~M. Yeomans and B.~Ladoux, \emph{{N}ature}, 2017,
  \textbf{544}, 212--216\relax
\mciteBstWouldAddEndPuncttrue
\mciteSetBstMidEndSepPunct{\mcitedefaultmidpunct}
{\mcitedefaultendpunct}{\mcitedefaultseppunct}\relax
\EndOfBibitem
\bibitem[Duclos \emph{et~al.}(2018)Duclos, Blanch~Mercader, Yashunsky,
  Salbreux, Joanny, Prost, and Silberzan]{Duclos2018}
G.~Duclos, C.~Blanch~Mercader, V.~Yashunsky, G.~Salbreux, J.~F. Joanny,
  J.~Prost and P.~Silberzan, \emph{{N}at. {P}hys.}, 2018, \textbf{14},
  728--732\relax
\mciteBstWouldAddEndPuncttrue
\mciteSetBstMidEndSepPunct{\mcitedefaultmidpunct}
{\mcitedefaultendpunct}{\mcitedefaultseppunct}\relax
\EndOfBibitem
\bibitem[Wensink \emph{et~al.}(2012)Wensink, Dunkel, Heidenreich, Drescher,
  Goldstein, L{\"o}wen, and Yeomans]{Wensink2012}
H.~H. Wensink, J.~Dunkel, S.~Heidenreich, K.~Drescher, R.~E. Goldstein,
  H.~L{\"o}wen and J.~M. Yeomans, \emph{{Proc}. {N}atl. {A}cad. {S}ci.
  {U}.{S}.{A}.}, 2012, \textbf{109}, 14308--14313\relax
\mciteBstWouldAddEndPuncttrue
\mciteSetBstMidEndSepPunct{\mcitedefaultmidpunct}
{\mcitedefaultendpunct}{\mcitedefaultseppunct}\relax
\EndOfBibitem
\bibitem[Dunkel \emph{et~al.}(2013)Dunkel, Heidenreich, Drescher, Wensink,
  B{\"a}r, and Goldstein]{Dunkel2013}
J.~Dunkel, S.~Heidenreich, K.~Drescher, H.~H. Wensink, M.~B{\"a}r and R.~E.
  Goldstein, \emph{{P}hys. {R}ev. {L}ett.}, 2013, \textbf{110}, 228102\relax
\mciteBstWouldAddEndPuncttrue
\mciteSetBstMidEndSepPunct{\mcitedefaultmidpunct}
{\mcitedefaultendpunct}{\mcitedefaultseppunct}\relax
\EndOfBibitem
\bibitem[Buhl \emph{et~al.}(2006)Buhl, Sumpter, Couzin, Hale, Despland, Miller,
  and Simpson]{Buhl2006}
J.~Buhl, D.~J.~T. Sumpter, I.~D. Couzin, J.~J. Hale, E.~Despland, E.~R. Miller
  and S.~J. Simpson, \emph{{S}cience}, 2006, \textbf{312}, 1402--1406\relax
\mciteBstWouldAddEndPuncttrue
\mciteSetBstMidEndSepPunct{\mcitedefaultmidpunct}
{\mcitedefaultendpunct}{\mcitedefaultseppunct}\relax
\EndOfBibitem
\bibitem[Lopez \emph{et~al.}(2012)Lopez, Gautrais, Couzin, and
  Theraulaz]{Lopez2012}
U.~Lopez, J.~Gautrais, I.~D. Couzin and G.~Theraulaz, \emph{{I}nterface
  {F}ocus}, 2012, \textbf{2}, 693--707\relax
\mciteBstWouldAddEndPuncttrue
\mciteSetBstMidEndSepPunct{\mcitedefaultmidpunct}
{\mcitedefaultendpunct}{\mcitedefaultseppunct}\relax
\EndOfBibitem
\bibitem[Harder and Cacciuto(2018)]{Harder2018}
J.~Harder and A.~Cacciuto, \emph{{P}hys. {R}ev. {E}}, 2018, \textbf{97},
  022603\relax
\mciteBstWouldAddEndPuncttrue
\mciteSetBstMidEndSepPunct{\mcitedefaultmidpunct}
{\mcitedefaultendpunct}{\mcitedefaultseppunct}\relax
\EndOfBibitem
\bibitem[Buttinoni \emph{et~al.}(2013)Buttinoni, Bialk{\'e}, K{\"u}mmel,
  L{\"o}wen, Bechinger, and Speck]{Buttinoni2013}
I.~Buttinoni, J.~Bialk{\'e}, F.~K{\"u}mmel, H.~L{\"o}wen, C.~Bechinger and
  T.~Speck, \emph{{P}hys. {R}ev. {L}ett.}, 2013, \textbf{110}, 238301\relax
\mciteBstWouldAddEndPuncttrue
\mciteSetBstMidEndSepPunct{\mcitedefaultmidpunct}
{\mcitedefaultendpunct}{\mcitedefaultseppunct}\relax
\EndOfBibitem
\bibitem[Narayan \emph{et~al.}(2007)Narayan, Ramaswamy, and Menon]{Narayan2007}
V.~Narayan, S.~Ramaswamy and N.~Menon, \emph{{S}cience}, 2007, \textbf{317},
  105--108\relax
\mciteBstWouldAddEndPuncttrue
\mciteSetBstMidEndSepPunct{\mcitedefaultmidpunct}
{\mcitedefaultendpunct}{\mcitedefaultseppunct}\relax
\EndOfBibitem
\bibitem[Marchetti \emph{et~al.}(2013)Marchetti, Joanny, Ramaswamy, Liverpool,
  Prost, Rao, and Simha]{Marchetti2013}
M.~C. Marchetti, J.~F. Joanny, S.~Ramaswamy, T.~B. Liverpool, J.~Prost, M.~Rao
  and R.~A. Simha, \emph{{R}ev. {M}od. {P}hys.}, 2013, \textbf{85},
  1143--1189\relax
\mciteBstWouldAddEndPuncttrue
\mciteSetBstMidEndSepPunct{\mcitedefaultmidpunct}
{\mcitedefaultendpunct}{\mcitedefaultseppunct}\relax
\EndOfBibitem
\bibitem[Koch and Subramanian(2011)]{Koch2011}
D.~L. Koch and G.~Subramanian, \emph{{A}nnu. {R}ev. {F}luid {M}ech.}, 2011,
  \textbf{43}, 637--659\relax
\mciteBstWouldAddEndPuncttrue
\mciteSetBstMidEndSepPunct{\mcitedefaultmidpunct}
{\mcitedefaultendpunct}{\mcitedefaultseppunct}\relax
\EndOfBibitem
\bibitem[Doostmohammadi \emph{et~al.}(2018)Doostmohammadi, Ign{\'e}s~Mullol,
  Yeomans, and Sagu{\'e}s]{Doostmohammadi2018}
A.~Doostmohammadi, J.~Ign{\'e}s~Mullol, J.~M. Yeomans and F.~Sagu{\'e}s,
  \emph{{N}at. {C}ommun.}, 2018, \textbf{9}, 3246\relax
\mciteBstWouldAddEndPuncttrue
\mciteSetBstMidEndSepPunct{\mcitedefaultmidpunct}
{\mcitedefaultendpunct}{\mcitedefaultseppunct}\relax
\EndOfBibitem
\bibitem[Ramaswamy(2010)]{Ramaswamy2010}
S.~Ramaswamy, \emph{{A}nnu. {R}ev. {C}ondens. {M}atter {P}hys.}, 2010,
  \textbf{1}, 323--345\relax
\mciteBstWouldAddEndPuncttrue
\mciteSetBstMidEndSepPunct{\mcitedefaultmidpunct}
{\mcitedefaultendpunct}{\mcitedefaultseppunct}\relax
\EndOfBibitem
\bibitem[Opathalage \emph{et~al.}(2018)Opathalage, Norton, Juniper, Aghvami,
  Langeslay, Fraden, and Dogic]{Opathalage2018}
A.~Opathalage, M.~M. Norton, M.~P. Juniper, S.~A. Aghvami, B.~Langeslay,
  S.~Fraden and Z.~Dogic, \emph{arXiv preprint arXiv:1810.09032}, 2018\relax
\mciteBstWouldAddEndPuncttrue
\mciteSetBstMidEndSepPunct{\mcitedefaultmidpunct}
{\mcitedefaultendpunct}{\mcitedefaultseppunct}\relax
\EndOfBibitem
\bibitem[Wu \emph{et~al.}(2017)Wu, Hishamunda, Chen, DeCamp, Chang,
  Fern{\'a}ndez~Nieves, Fraden, and Dogic]{Wu2017}
K.~T. Wu, J.~B. Hishamunda, D.~T.~N. Chen, S.~J. DeCamp, Y.~W. Chang,
  A.~Fern{\'a}ndez~Nieves, S.~Fraden and Z.~Dogic, \emph{{S}cience}, 2017,
  \textbf{355}, eaal1979\relax
\mciteBstWouldAddEndPuncttrue
\mciteSetBstMidEndSepPunct{\mcitedefaultmidpunct}
{\mcitedefaultendpunct}{\mcitedefaultseppunct}\relax
\EndOfBibitem
\bibitem[Sanchez \emph{et~al.}(2012)Sanchez, Chen, DeCamp, Heymann, and
  Dogic]{Sanchez2012}
T.~Sanchez, D.~T.~N. Chen, S.~J. DeCamp, M.~Heymann and Z.~Dogic,
  \emph{{N}ature}, 2012, \textbf{491}, 431--434\relax
\mciteBstWouldAddEndPuncttrue
\mciteSetBstMidEndSepPunct{\mcitedefaultmidpunct}
{\mcitedefaultendpunct}{\mcitedefaultseppunct}\relax
\EndOfBibitem
\bibitem[Volfson \emph{et~al.}(2008)Volfson, Cookson, Hasty, and
  Tsimring]{Volfson2008}
D.~Volfson, S.~Cookson, J.~Hasty and L.~S. Tsimring, \emph{{Proc}. {N}atl.
  {A}cad. {S}ci. {U}.{S}.{A}.}, 2008, \textbf{105}, 15346--15351\relax
\mciteBstWouldAddEndPuncttrue
\mciteSetBstMidEndSepPunct{\mcitedefaultmidpunct}
{\mcitedefaultendpunct}{\mcitedefaultseppunct}\relax
\EndOfBibitem
\bibitem[Doostmohammadi \emph{et~al.}(2016)Doostmohammadi, Adamer, Thampi, and
  Yeomans]{Doostmohammadi2016}
A.~Doostmohammadi, M.~F. Adamer, S.~P. Thampi and J.~M. Yeomans, \emph{{N}at.
  {C}ommun.}, 2016, \textbf{7}, 10557\relax
\mciteBstWouldAddEndPuncttrue
\mciteSetBstMidEndSepPunct{\mcitedefaultmidpunct}
{\mcitedefaultendpunct}{\mcitedefaultseppunct}\relax
\EndOfBibitem
\bibitem[Shi \emph{et~al.}(2014)Shi, Chat{\'e}, and Ma]{Shi2014}
X.~q. Shi, H.~Chat{\'e} and Y.~q. Ma, \emph{{N}ew {J}. {P}hys.}, 2014,
  \textbf{16}, 035003\relax
\mciteBstWouldAddEndPuncttrue
\mciteSetBstMidEndSepPunct{\mcitedefaultmidpunct}
{\mcitedefaultendpunct}{\mcitedefaultseppunct}\relax
\EndOfBibitem
\bibitem[Kemkemer \emph{et~al.}(2000)Kemkemer, Teichgr{\"a}ber,
  Schrank~Kaufmann, Kaufmann, and Gruler]{Kemkemer2000}
R.~Kemkemer, V.~Teichgr{\"a}ber, S.~Schrank~Kaufmann, D.~Kaufmann and
  H.~Gruler, \emph{{E}ur. {P}hys. {J}. {E}}, 2000, \textbf{3}, 101--110\relax
\mciteBstWouldAddEndPuncttrue
\mciteSetBstMidEndSepPunct{\mcitedefaultmidpunct}
{\mcitedefaultendpunct}{\mcitedefaultseppunct}\relax
\EndOfBibitem
\bibitem[Duclos \emph{et~al.}(2014)Duclos, Garcia, Yevick, and
  Silberzan]{Duclos2014}
G.~Duclos, S.~Garcia, H.~G. Yevick and P.~Silberzan, \emph{{S}oft {M}atter},
  2014, \textbf{10}, 2346--2353\relax
\mciteBstWouldAddEndPuncttrue
\mciteSetBstMidEndSepPunct{\mcitedefaultmidpunct}
{\mcitedefaultendpunct}{\mcitedefaultseppunct}\relax
\EndOfBibitem
\bibitem[Kawaguchi \emph{et~al.}(2017)Kawaguchi, Kageyama, and
  Sano]{Kawaguchi2017}
K.~Kawaguchi, R.~Kageyama and M.~Sano, \emph{{N}ature}, 2017, \textbf{545},
  327--331\relax
\mciteBstWouldAddEndPuncttrue
\mciteSetBstMidEndSepPunct{\mcitedefaultmidpunct}
{\mcitedefaultendpunct}{\mcitedefaultseppunct}\relax
\EndOfBibitem
\bibitem[Voituriez \emph{et~al.}(2005)Voituriez, Joanny, and
  Prost]{Voituriez2005}
R.~Voituriez, J.~F. Joanny and J.~Prost, \emph{{E}urophys. {L}ett.}, 2005,
  \textbf{70}, 404--410\relax
\mciteBstWouldAddEndPuncttrue
\mciteSetBstMidEndSepPunct{\mcitedefaultmidpunct}
{\mcitedefaultendpunct}{\mcitedefaultseppunct}\relax
\EndOfBibitem
\bibitem[Shendruk \emph{et~al.}(2017)Shendruk, Doostmohammadi, Thijssen, and
  Yeomans]{Shendruk2017}
T.~N. Shendruk, A.~Doostmohammadi, K.~Thijssen and J.~M. Yeomans, \emph{{S}oft
  {M}atter}, 2017, \textbf{13}, 3853--3862\relax
\mciteBstWouldAddEndPuncttrue
\mciteSetBstMidEndSepPunct{\mcitedefaultmidpunct}
{\mcitedefaultendpunct}{\mcitedefaultseppunct}\relax
\EndOfBibitem
\bibitem[Thampi \emph{et~al.}(2013)Thampi, Golestanian, and
  Yeomans]{Thampi2013}
S.~P. Thampi, R.~Golestanian and J.~M. Yeomans, \emph{{P}hys. {R}ev. {L}ett.},
  2013, \textbf{111}, 118101\relax
\mciteBstWouldAddEndPuncttrue
\mciteSetBstMidEndSepPunct{\mcitedefaultmidpunct}
{\mcitedefaultendpunct}{\mcitedefaultseppunct}\relax
\EndOfBibitem
\bibitem[Giomi \emph{et~al.}(2013)Giomi, Bowick, Ma, and Marchetti]{Giomi2013}
L.~Giomi, M.~J. Bowick, X.~Ma and M.~C. Marchetti, \emph{{P}hys. {R}ev.
  {L}ett.}, 2013, \textbf{110}, 228101\relax
\mciteBstWouldAddEndPuncttrue
\mciteSetBstMidEndSepPunct{\mcitedefaultmidpunct}
{\mcitedefaultendpunct}{\mcitedefaultseppunct}\relax
\EndOfBibitem
\bibitem[Thampi \emph{et~al.}(2014)Thampi, Golestanian, and
  Yeomans]{Thampi2014}
S.~P. Thampi, R.~Golestanian and J.~M. Yeomans, \emph{{P}hys. {R}ev. {E}},
  2014, \textbf{90}, 062307\relax
\mciteBstWouldAddEndPuncttrue
\mciteSetBstMidEndSepPunct{\mcitedefaultmidpunct}
{\mcitedefaultendpunct}{\mcitedefaultseppunct}\relax
\EndOfBibitem
\bibitem[Doostmohammadi \emph{et~al.}(2017)Doostmohammadi, Shendruk, Thijssen,
  and Yeomans]{Doostmohammadi2017}
A.~Doostmohammadi, T.~N. Shendruk, K.~Thijssen and J.~M. Yeomans, \emph{{N}at.
  {C}ommun.}, 2017, \textbf{8}, 15326\relax
\mciteBstWouldAddEndPuncttrue
\mciteSetBstMidEndSepPunct{\mcitedefaultmidpunct}
{\mcitedefaultendpunct}{\mcitedefaultseppunct}\relax
\EndOfBibitem
\bibitem[de~Gennes and Prost(1995)]{P1995}
P.~G. de~Gennes and J.~Prost, \emph{{T}he physics of liquid crystals}, Oxford
  University Press, UK, 1995\relax
\mciteBstWouldAddEndPuncttrue
\mciteSetBstMidEndSepPunct{\mcitedefaultmidpunct}
{\mcitedefaultendpunct}{\mcitedefaultseppunct}\relax
\EndOfBibitem
\bibitem[Yeomans(2016)]{Yeomans2016}
J.~M. Yeomans, \emph{in {S}oft {M}atter {S}elf-{A}ssembly}, IOS Press,
  Amsterdam, 2016, pp. 383--415\relax
\mciteBstWouldAddEndPuncttrue
\mciteSetBstMidEndSepPunct{\mcitedefaultmidpunct}
{\mcitedefaultendpunct}{\mcitedefaultseppunct}\relax
\EndOfBibitem
\bibitem[Beris and Edwards(1994)]{Beris1994}
A.~N. Beris and B.~J. Edwards, \emph{{T}hermodynamics of Flowing Systems: with
  Internal Microstructure}, Oxford University Press, UK, 1994\relax
\mciteBstWouldAddEndPuncttrue
\mciteSetBstMidEndSepPunct{\mcitedefaultmidpunct}
{\mcitedefaultendpunct}{\mcitedefaultseppunct}\relax
\EndOfBibitem
\bibitem[Batchelor(2000)]{Batchelor2000}
G.~Batchelor, \emph{{A}n {I}ntroduction to {F}luid {D}ynamics}, Cambridge
  University Press, UK, illustrated, reprint edn., 2000\relax
\mciteBstWouldAddEndPuncttrue
\mciteSetBstMidEndSepPunct{\mcitedefaultmidpunct}
{\mcitedefaultendpunct}{\mcitedefaultseppunct}\relax
\EndOfBibitem
\bibitem[Simha and Ramaswamy(2002)]{Simha2002}
R.~A. Simha and S.~Ramaswamy, \emph{{P}hys. {R}ev. {L}ett.}, 2002, \textbf{89},
  058101\relax
\mciteBstWouldAddEndPuncttrue
\mciteSetBstMidEndSepPunct{\mcitedefaultmidpunct}
{\mcitedefaultendpunct}{\mcitedefaultseppunct}\relax
\EndOfBibitem
\bibitem[Edwards and Yeomans(2009)]{Edwards2009}
S.~A. Edwards and J.~M. Yeomans, \emph{{E}urophys. {L}ett.}, 2009, \textbf{85},
  18008\relax
\mciteBstWouldAddEndPuncttrue
\mciteSetBstMidEndSepPunct{\mcitedefaultmidpunct}
{\mcitedefaultendpunct}{\mcitedefaultseppunct}\relax
\EndOfBibitem
\bibitem[Denniston \emph{et~al.}(2002)Denniston, T{\'o}th, and
  Yeomans]{Denniston2002}
C.~Denniston, G.~T{\'o}th and J.~M. Yeomans, \emph{{J}. {S}tat. {P}hys.}, 2002,
  \textbf{107}, 187--202\relax
\mciteBstWouldAddEndPuncttrue
\mciteSetBstMidEndSepPunct{\mcitedefaultmidpunct}
{\mcitedefaultendpunct}{\mcitedefaultseppunct}\relax
\EndOfBibitem
\bibitem[Frenkel(2014)]{Frenkel2014}
D.~Frenkel, \emph{{N}at. {M}ater.}, 2014, \textbf{14}, 9\relax
\mciteBstWouldAddEndPuncttrue
\mciteSetBstMidEndSepPunct{\mcitedefaultmidpunct}
{\mcitedefaultendpunct}{\mcitedefaultseppunct}\relax
\EndOfBibitem
\bibitem[Thampi \emph{et~al.}(2015)Thampi, Doostmohammadi, Golestanian, and
  Yeomans]{Thampi2015}
S.~P. Thampi, A.~Doostmohammadi, R.~Golestanian and J.~M. Yeomans,
  \emph{{E}urophys. {L}ett.}, 2015, \textbf{112}, 28004\relax
\mciteBstWouldAddEndPuncttrue
\mciteSetBstMidEndSepPunct{\mcitedefaultmidpunct}
{\mcitedefaultendpunct}{\mcitedefaultseppunct}\relax
\EndOfBibitem
\bibitem[Guillamat \emph{et~al.}(2017)Guillamat, Ign{\'e}s~Mullol, and
  Sagu{\'e}s]{Guillamat2017}
P.~Guillamat, J.~Ign{\'e}s~Mullol and F.~Sagu{\'e}s, \emph{{N}at. {C}ommun.},
  2017, \textbf{8}, 564\relax
\mciteBstWouldAddEndPuncttrue
\mciteSetBstMidEndSepPunct{\mcitedefaultmidpunct}
{\mcitedefaultendpunct}{\mcitedefaultseppunct}\relax
\EndOfBibitem
\bibitem[Giomi(2015)]{Giomi2015}
L.~Giomi, \emph{{P}hys. {R}ev. {X}}, 2015, \textbf{5}, 031003\relax
\mciteBstWouldAddEndPuncttrue
\mciteSetBstMidEndSepPunct{\mcitedefaultmidpunct}
{\mcitedefaultendpunct}{\mcitedefaultseppunct}\relax
\EndOfBibitem
\bibitem[Marenduzzo \emph{et~al.}(2007)Marenduzzo, Orlandini, Cates, and
  Yeomans]{Marenduzzo2007}
D.~Marenduzzo, E.~Orlandini, M.~E. Cates and J.~M. Yeomans, \emph{{P}hys.
  {R}ev. {E}: {Stat}., {N}onlinear, {S}oft {M}atter {P}hys.}, 2007,
  \textbf{76}, 031921\relax
\mciteBstWouldAddEndPuncttrue
\mciteSetBstMidEndSepPunct{\mcitedefaultmidpunct}
{\mcitedefaultendpunct}{\mcitedefaultseppunct}\relax
\EndOfBibitem
\bibitem[Nier \emph{et~al.}(2016)Nier, Jain, Lim, Ishihara, Ladoux, and
  Marcq]{Nier2016}
V.~Nier, S.~Jain, C.~T. Lim, S.~Ishihara, B.~Ladoux and P.~Marcq,
  \emph{{B}iophys. {J}.}, 2016, \textbf{110}, 1625--1635\relax
\mciteBstWouldAddEndPuncttrue
\mciteSetBstMidEndSepPunct{\mcitedefaultmidpunct}
{\mcitedefaultendpunct}{\mcitedefaultseppunct}\relax
\EndOfBibitem
\bibitem[Wioland \emph{et~al.}(2016)Wioland, Lushi, and Goldstein]{Wioland2016}
H.~Wioland, E.~Lushi and R.~E. Goldstein, \emph{{N}ew {J}. {P}hys.}, 2016,
  \textbf{18}, 075002\relax
\mciteBstWouldAddEndPuncttrue
\mciteSetBstMidEndSepPunct{\mcitedefaultmidpunct}
{\mcitedefaultendpunct}{\mcitedefaultseppunct}\relax
\EndOfBibitem
\end{mcitethebibliography}
\bibliographystyle{rsc.bst} %the RSC's .bst file

\end{document}